
\catcode`\@=11					



\font\fiverm=cmr5				
\font\fivemi=cmmi5				
\font\fivesy=cmsy5				
\font\fivebf=cmbx5				

\skewchar\fivemi='177
\skewchar\fivesy='60


\font\sixrm=cmr6				
\font\sixi=cmmi6				
\font\sixsy=cmsy6				
\font\sixbf=cmbx6				

\skewchar\sixi='177
\skewchar\sixsy='60


\font\sevenrm=cmr7				
\font\seveni=cmmi7				
\font\sevensy=cmsy7				
\font\sevenit=cmti7				
\font\sevenbf=cmbx7				

\skewchar\seveni='177
\skewchar\sevensy='60


\font\eightrm=cmr8				
\font\eighti=cmmi8				
\font\eightsy=cmsy8				
\font\eightit=cmti8				
\font\eightbf=cmbx8				

\skewchar\eighti='177
\skewchar\eightsy='60


\font\ninei=cmmi9
\font\ninesy=cmsy9

\skewchar\ninei='177
\skewchar\ninesy='60


\font\tenrm=cmr10				
\font\teni=cmmi10				
\font\tensy=cmsy10				
\font\tenex=cmex10				
\font\tenit=cmti10				
\font\tensl=cmsl10				
\font\tenbf=cmbx10				
\font\tentt=cmtt10				
\font\tenss=cmss10				
\font\tensc=cmcsc10				
\font\tenbi=cmmib10				

\skewchar\teni='177
\skewchar\tenbi='177
\skewchar\tensy='60

\def\tenpoint{\ifmmode\err@badsizechange\else
	\textfont0=\tenrm \scriptfont0=\sevenrm \scriptscriptfont0=\fiverm
	\textfont1=\teni  \scriptfont1=\seveni  \scriptscriptfont1=\fivemi
	\textfont2=\tensy \scriptfont2=\sevensy \scriptscriptfont2=\fivesy
	\textfont3=\tenex \scriptfont3=\tenex   \scriptscriptfont3=\tenex
	\textfont4=\tenit \scriptfont4=\sevenit \scriptscriptfont4=\sevenit
	\textfont5=\tensl
	\textfont6=\tenbf \scriptfont6=\sevenbf \scriptscriptfont6=\fivebf
	\textfont7=\tentt
	\textfont8=\tenbi \scriptfont8=\seveni  \scriptscriptfont8=\fivemi
	\def\rm{\tenrm\fam=0 }%
	\def\it{\tenit\fam=4 }%
	\def\sl{\tensl\fam=5 }%
	\def\bf{\tenbf\fam=6 }%
	\def\tt{\tentt\fam=7 }%
	\def\ss{\tenss}%
	\def\sc{\tensc}%
	\def\bmit{\fam=8 }%
	\rm\setparameters\setbaselines\fi}


\font\twelverm=cmr12				
\font\twelvei=cmmi12				
\font\twelvesy=cmsy10	scaled\magstep1		
\font\twelveex=cmex10	scaled\magstep1		
\font\twelveit=cmti12				
\font\twelvesl=cmsl12				
\font\twelvebf=cmbx12				
\font\twelvett=cmtt12				
\font\twelvess=cmss12				
\font\twelvesc=cmcsc10	scaled\magstep1		
\font\twelvebi=cmmib10	scaled\magstep1		

\skewchar\twelvei='177
\skewchar\twelvebi='177
\skewchar\twelvesy='60

\def\twelvepoint{\ifmmode\err@badsizechange\else
	\textfont0=\twelverm \scriptfont0=\eightrm \scriptscriptfont0=\sixrm
	\textfont1=\twelvei  \scriptfont1=\eighti  \scriptscriptfont1=\sixi
	\textfont2=\twelvesy \scriptfont2=\eightsy \scriptscriptfont2=\sixsy
	\textfont3=\twelveex \scriptfont3=\tenex   \scriptscriptfont3=\tenex
	\textfont4=\twelveit \scriptfont4=\eightit \scriptscriptfont4=\sevenit
	\textfont5=\twelvesl
	\textfont6=\twelvebf \scriptfont6=\eightbf \scriptscriptfont6=\sixbf
	\textfont7=\twelvett
	\textfont8=\twelvebi \scriptfont8=\eighti  \scriptscriptfont8=\sixi
	\def\rm{\twelverm\fam=0 }%
	\def\it{\twelveit\fam=4 }%
	\def\sl{\twelvesl\fam=5 }%
	\def\bf{\twelvebf\fam=6 }%
	\def\tt{\twelvett\fam=7 }%
	\def\ss{\twelvess}%
	\def\sc{\twelvesc}%
	\def\bmit{\fam=8 }%
	\rm\setparameters\setbaselines\fi}


\font\fourteenrm=cmr10	scaled\magstep2		
\font\fourteeni=cmmi10	scaled\magstep2		
\font\fourteensy=cmsy10	scaled\magstep2		
\font\fourteenex=cmex10	scaled\magstep2		
\font\fourteenit=cmti10	scaled\magstep2		
\font\fourteensl=cmsl10	scaled\magstep2		
\font\fourteenbf=cmbx10	scaled\magstep2		
\font\fourteentt=cmtt10	scaled\magstep2		
\font\fourteenss=cmss10	scaled\magstep2		
\font\fourteensc=cmcsc10 scaled\magstep2	
\font\fourteenbi=cmmib10 scaled\magstep2	

\skewchar\fourteeni='177
\skewchar\fourteenbi='177
\skewchar\fourteensy='60

\def\fourteenpoint{\ifmmode\err@badsizechange\else
	\textfont0=\fourteenrm \scriptfont0=\tenrm \scriptscriptfont0=\sevenrm
	\textfont1=\fourteeni  \scriptfont1=\teni  \scriptscriptfont1=\seveni
	\textfont2=\fourteensy \scriptfont2=\tensy \scriptscriptfont2=\sevensy
	\textfont3=\fourteenex \scriptfont3=\tenex \scriptscriptfont3=\tenex
	\textfont4=\fourteenit \scriptfont4=\tenit \scriptscriptfont4=\sevenit
	\textfont5=\fourteensl
	\textfont6=\fourteenbf \scriptfont6=\tenbf \scriptscriptfont6=\sevenbf
	\textfont7=\fourteentt
	\textfont8=\fourteenbi \scriptfont8=\tenbi \scriptscriptfont8=\seveni
	\def\rm{\fourteenrm\fam=0 }%
	\def\it{\fourteenit\fam=4 }%
	\def\sl{\fourteensl\fam=5 }%
	\def\bf{\fourteenbf\fam=6 }%
	\def\tt{\fourteentt\fam=7}%
	\def\ss{\fourteenss}%
	\def\sc{\fourteensc}%
	\def\bmit{\fam=8 }%
	\rm\setparameters\setbaselines\fi}


\font\seventeenrm=cmr10 scaled\magstep3		


\newdimen\rp@
\newcount\@basestretchnum
\newskip\@baseskip
\newskip\headskip
\newskip\footskip


\def\setparameters{\rp@=.1em
	\headskip=24\rp@
	\footskip=\headskip
	\delimitershortfall=5\rp@
	\nulldelimiterspace=1.2\rp@
	\scriptspace=0.5\rp@
	\abovedisplayskip=10\rp@ plus3\rp@ minus5\rp@
	\belowdisplayskip=10\rp@ plus3\rp@ minus5\rp@
	\abovedisplayshortskip=5\rp@ plus2\rp@ minus4\rp@
	\belowdisplayshortskip=10\rp@ plus3\rp@ minus5\rp@
	\normallineskip=\rp@
	\lineskip=\normallineskip
	\normallineskiplimit=0pt
	\lineskiplimit=\normallineskiplimit
	\jot=3\rp@
	\setbox0=\hbox{\the\textfont3 B}\p@renwd=\wd0
	\skip\footins=12\rp@ plus3\rp@ minus3\rp@
	\skip\topins=0pt plus0pt minus0pt}


\def\setbaselines{\maxdepth=4\rp@\baselinestretch=\@basestretchnum}


\def\baselinestretch{\afterassignment\@basestretch\@basestretchnum}
\def\@basestretch{%
	\@baseskip=12\rp@ \divide\@baseskip by1000
	\normalbaselineskip=\@basestretchnum\@baseskip
	\baselineskip=\normalbaselineskip
	\bigskipamount=\the\baselineskip
		plus.25\baselineskip minus.25\baselineskip
	\medskipamount=.5\baselineskip
		plus.125\baselineskip minus.125\baselineskip
	\smallskipamount=.25\baselineskip
		plus.0625\baselineskip minus.0625\baselineskip
	\setbox\strutbox=\hbox{\vrule height.708\baselineskip
		depth.292\baselineskip width0pt }}



\def\makeheadline{\vbox to0pt{\baselinestretch=1000
	\vskip-\headskip \vskip1.5pt
	\line{\vbox to\ht\strutbox{}\the\headline}\vss}\nointerlineskip}

\def\makefootline{\baselineskip=\footskip\line{\the\footline}}

\def\big#1{{\hbox{$\left#1\vbox to8.5\rp@ {}\right.\n@space$}}}
\def\Big#1{{\hbox{$\left#1\vbox to11.5\rp@ {}\right.\n@space$}}}
\def\bigg#1{{\hbox{$\left#1\vbox to14.5\rp@ {}\right.\n@space$}}}
\def\Bigg#1{{\hbox{$\left#1\vbox to17.5\rp@ {}\right.\n@space$}}}


\mathchardef\alpha="710B
\mathchardef\beta="710C
\mathchardef\gamma="710D
\mathchardef\delta="710E
\mathchardef\epsilon="710F
\mathchardef\zeta="7110
\mathchardef\eta="7111
\mathchardef\theta="7112
\mathchardef\iota="7113
\mathchardef\kappa="7114
\mathchardef\lambda="7115
\mathchardef\mu="7116
\mathchardef\nu="7117
\mathchardef\xi="7118
\mathchardef\pi="7119
\mathchardef\rho="711A
\mathchardef\sigma="711B
\mathchardef\tau="711C
\mathchardef\upsilon="711D
\mathchardef\phi="711E
\mathchardef\chi="711F
\mathchardef\psi="7120
\mathchardef\omega="7121
\mathchardef\varepsilon="7122
\mathchardef\vartheta="7123
\mathchardef\varpi="7124
\mathchardef\varrho="7125
\mathchardef\varsigma="7126
\mathchardef\varphi="7127
\mathchardef\imath="717B
\mathchardef\jmath="717C
\mathchardef\ell="7160
\mathchardef\wp="717D
\mathchardef\partial="7140
\mathchardef\flat="715B
\mathchardef\natural="715C
\mathchardef\sharp="715D


\def\err@badsizechange{%
	\immediate\write16{--> Size change not allowed in math mode, ignored}}

\baselinestretch=1000
\tenpoint

\catcode`\@=12					
\catcode`\@=11
\expandafter\ifx\csname @iasmacros\endcsname\relax
	\global\let\@iasmacros=\par
\else	\endinput
\fi
\catcode`\@=12


\def\rmb{\seventeenrm}

\def\itb{icmsy8}


\def\halfspace{\baselineskip=1.5\normalbaselineskip}
\def\doublespace{\baselineskip=2\normalbaselineskip}


\def\AB{\bigskip\parindent=40pt
        \centerline{\bf ABSTRACT}\medskip\halfspace\narrower}
\def\AE{\bigskip\nonarrower\doublespace}
\def\nonarrower{\advance\leftskip by-\parindent
	\advance\rightskip by-\parindent}


\def\boxit#1{\vbox{\hrule\hbox{\vrule\kern3pt
	\vbox{\kern3pt#1\kern3pt}\kern3pt\vrule}\hrule}}

\def\hence{\leavevmode\hbox{\bf .\raise5.5pt\hbox{.}.} }

\def\dalemb#1#2{{\vbox{\hrule height.#2pt
	\hbox{\vrule width.#2pt height#1pt \kern#1pt \vrule width.#2pt}
	\hrule height.#2pt}}}
\def\gtorder{\mathrel{\raise.3ex\hbox{$>$}\mkern-14mu
             \lower0.6ex\hbox{$\sim$}}}
\def\ltorder{\mathrel{\raise.3ex\hbox{$<$}\mkern-14mu
             \lower0.6ex\hbox{$\sim$}}}

\newdimen\fullhsize
\newbox\leftcolumn
\def\twoup{\hoffset=-.5in \voffset=-.25in
  \hsize=4.75in \fullhsize=10in \vsize=6.9in
  \def\fullline{\hbox to\fullhsize}
  \let\lr=L
  \output={\if L\lr
        \global\setbox\leftcolumn=\columnbox\global\let\lr=R \advancepageno
      \else \doubleformat \global\let\lr=L\fi
    \ifnum\outputpenalty>-20000 \else\dosupereject\fi}
  \def\doubleformat{\shipout\vbox{
    \fullline{\box\leftcolumn\hfil\columnbox}\advancepageno}}
  \def\columnbox{\leftline{\vbox{\makeheadline\pagebody\makefootline}}}
  \tolerance=1000 }

\twelvepoint
\doublespace
\font\itb=cmbxti10 scaled\magstep1
{{
\rightline{IASSNS-HEP-95/66}
\rightline{~~~SECOND REVISION December, 1995}
\bigskip\bigskip
\centerline{\rmb  Generalized Quantum Dynamics as Pre-Quantum Mechanics}
\medskip
\centerline{\itb Stephen L. Adler
}
\centerline{\bf Institute for Advanced Study}
\centerline{\bf Princeton, NJ 08540}
\medskip
\centerline{\itb Andrew C. Millard}
\centerline{\bf Department of Physics, Jadwin Laboratory}
\centerline{\bf Princeton University, Princeton, NJ 08544}
\medskip
\bigskip
\bigskip

\vfill\eject
\pageno=2
\AB
We address the issue of when generalized quantum dynamics, which is a
classical symplectic dynamics for noncommuting operator phase space
variables based on a graded
total trace Hamiltonian ${\bf H}$, reduces to Heisenberg picture complex
quantum mechanics.  We begin by showing that when ${\bf H}={\bf Tr}
H$, with $H$ a Weyl ordered operator Hamiltonian, then the
generalized quantum dynamics operator equations of motion agree with those
obtained from $H$ in the Heisenberg picture by using canonical
commutation relations.  The  remainder of the paper is devoted to a study
of how an effective canonical algebra can arise, without this condition
simply being imposed by fiat on the operator initial values.
We first show that for any total trace Hamiltonian which involves no
noncommutative constants, there is a conserved anti--self--adjoint
operator $\tilde C$ with a structure
which is closely related to the canonical commutator algebra.
We study the canonical
transformations of generalized quantum dynamics, and show that $\tilde C$ is
a canonical invariant, as is the operator phase space volume element. The
latter result is a generalization of
Liouville's theorem, and permits the application
of statistical mechanical methods to determine the canonical ensemble
governing the equilibrium distribution of operator initial values.
We give arguments based on a Ward identity
analogous to the equipartition theorem of classical statistical mechanics,
suggesting that statistical ensemble averages of Weyl ordered polynomials in
the operator phase space
variables correspond to the Wightman functions of a unitary
complex quantum mechanics, with a conserved operator Hamiltonian and with
the standard canonical commutation relations obeyed by Weyl
ordered operator strings.  Thus there is a well--defined sense in which
complex quantum field theory can emerge as a statistical approximation to
an underlying generalized quantum dynamics.

\AE
\bigskip\bigskip
\vfill\eject
\pageno=3
\centerline{{\bf 1.~~Introduction and Brief Review of}}
\centerline{{\bf Generalized Quantum Dynamics}}

In several recent publications one of us (S.L.A.) formulated [1, 2] and
with collaborators elaborated [3, 4] an
operator dynamics, called {\it generalized quantum dynamics}, which gives
a symplectic dynamics for general noncommutative degrees of freedom.  This
permits the direct derivation of equations of motion for field operators,
dispensing with the conventional canonical procedure of ``quantizing'' a
classical theory.  Although we observed that generalized quantum dynamics
in a complex Hilbert space is compatible with canonical quantization, the
precise connection between the two formalisms was not established,
and it is this issue which we address in the present paper.  We will not
in fact find it necessary to restrict ourselves to complex Hilbert space,
and the derivations and conclusions given here apply (with some specific
differences which we discuss) in quaternionic Hilbert space and real
Hilbert space as well.

Generalized quantum dynamics can be given in either Lagrangian or
Hamiltonian form, and for brevity we review only the Hamiltonian formalism,
since this is what we will need.  We shall assume an underlying Hilbert
space which is the  direct sum of bosonic and fermionic subspaces, and a
grading operator $(-1)^F$ with eigenvalue $1 (-1)$ for states in the bosonic
(fermionic) subspace.  For a general operator ${\cal O}$, we define the
graded trace operation ${\bf Tr} {\cal O}$ by
$$\eqalign{
{\bf Tr}{\cal O} =& {\rm Re}{\rm Tr} (-1)^F {\cal O}
={\rm Re} \sum_n \langle n|(-1)^F {\cal O}  |n \rangle ~~~\cr
=&{\rm Re} \sum_{n,B} \langle n|{\cal O}  |n \rangle-
{\rm Re} \sum_{n,F} \langle n|{\cal O}  |n \rangle ~~~,\cr
}\eqno(1)$$
with the subscripts $B,F$ on the sums indicating summations over bosonic
and fermionic states, respectively.
We call operators {\it bosonic} if they commute with $(-1)^F$ and
{\it fermionic} if they anticommute with $(-1)^F$.  Given sufficient
convergence,
it is then easy to see that ${\bf Tr} {\cal O}$ vanishes if ${\cal O}$ is
fermionic, and that {\bf Tr} obeys the {\it cyclic property}
$${\bf Tr} {\cal O}_{(1)} {\cal O}_{(2)} =
\pm {\bf Tr} {\cal O}_{(2)} {\cal O}_{(1)} ~~~, \eqno(2)$$
with the $+(-)$ sign holding when ${\cal O}_{(1)}$ and ${\cal} O_{(2)}$ are
both bosonic (fermionic).

Let now $\{ q_r(t) \},\{ p_r(t) \}, ~~~r=1,...,N$ be a set of operator phase
space variables.  For each $r$, we assume that $q_r$ and $p_r$ are
either both
bosonic or both fermionic, but we make no {\it a priori} assumption about
the commutativity of the phase space variables with one another.  Letting
$A[ \{q_r\},\{p_r\}]$  be a polynomial (or Laurent expandable) operator
function
of the phase space variables, we define the real number--valued
{\it total trace functional}
{\bf A} by
$${\bf A} [ \{q_r\},\{p_r\}] = {\bf Tr} A[ \{q_r\},\{p_r \}] ~~~.
\eqno(3)$$
Although noncommutativity of the phase space variables prevents us from
simply differentiating  $A$ with respect to them, we can use the
cyclic property of {\bf Tr} to define derivatives of {\bf A} by forming
$\delta {\bf A}$ and cyclically reordering all the operator variations
$\delta q_r,\delta p_r$
to the right, giving the fundamental definition
$$\delta {\bf A}= {\bf Tr} \sum_r \left( {\delta {\bf A} \over \delta q_r}
\delta q_r + {\delta {\bf A} \over \delta p_r} \delta p_r \right)~~~,
\eqno(4)$$
in which $\delta {\bf A} / \delta q_r$ and
$\delta {\bf A} / \delta p_r$ are themselves operators.

Let us now introduce an operator Hamiltonian $H[ \{q_r\},\{p_r\}]$
and a corresponding total trace Hamiltonian ${\bf H} = {\bf Tr} H$,
which generates the dynamics of the phase space variables via the
operator Hamilton equations
$${\delta {\bf H} \over \delta q_r} = - \dot p_r~~,~~~
{\delta {\bf H} \over \delta p_r} = \epsilon_r  \dot q_r~~~, \eqno(5)$$
with $\epsilon_r= 1 (-1)$ according to whether $q_r$ and $p_r$ are
bosonic (fermionic), and with the dot denoting the time derivative.
(As discussed in Refs.~[1, 2], this Hamiltonian
formulation can be derived from a total trace Lagrangian action principle,
in strict analogy with standard derivations of classical mechanics.)
If ${\bf A}[ \{q_r\},\{p_r\}, t]$ is an arbitrary total trace functional
which can have an explicit time dependence, as well as an implicit time
dependence through its dependence on the phase space variables, then a
simple application of Eqs.~(1--5) shows that the total time derivative
of {\bf A} is given by
$${d{\bf A} \over dt}= {\partial {\bf A} \over \partial t}
+ \{ {\bf A} , {\bf H} \}~~~, \eqno(6a)$$
where we have denoted by $\{{\bf A}, {\bf B} \}$ the {\it generalized
Poisson bracket} defined by
$$\{ {\bf A}, {\bf B} \} = {\bf Tr} \sum_{r=1}^N \epsilon_r
\left( {\delta {\bf A} \over \delta q_r} {\delta {\bf B} \over \delta p_r}
- {\delta {\bf B} \over \delta q_r} {\delta {\bf A} \over \delta p_r} \right)
{}~~~~. \eqno(6b)$$
Since the generalized Poisson bracket is antisymmetric
in its arguments, by taking {\bf A} to be the total trace Hamiltonian {\bf H},
which has no explicit time dependence,  we learn from Eq.~(6a) that
{\bf H} is a constant of the motion.

In addition to its antisymmetry, the generalized Poisson bracket can also
be shown [3] to satisfy the Jacobi identity
$$0=\{ {\bf A}, \{ {\bf B}, {\bf C} \} \}
+ \{ {\bf C}, \{ {\bf A}, {\bf B} \} \}
+\{ {\bf B}, \{ {\bf C}, {\bf A} \} \} ~~~. \eqno(7)$$
As a consequence, the phase space flows in generalized quantum dynamics
exhibit many features [4] analogous to those of ordinary classical mechanics.
In order to exhibit the symplectic structure of the generalized Poisson
bracket, we use the cyclic property of {\bf Tr} to
rewrite Eq.~(6b) as
$$\{ {\bf A}, {\bf B} \} = {\bf Tr} \left[ \sum_{r,B}
\left( {\delta {\bf A} \over \delta q_r} {\delta {\bf B} \over \delta p_r}
- {\delta {\bf A} \over \delta p_r} {\delta {\bf B} \over \delta q_r} \right)
- \sum_{r,F}
\left( {\delta {\bf A} \over \delta q_r} {\delta {\bf B} \over \delta p_r}
+ {\delta {\bf A} \over \delta p_r} {\delta {\bf B} \over \delta q_r} \right)
\right]
{}~~~~, \eqno(8)$$
with the subscripts $B,F$ on the sums respectively indicating
summations over the bosonic and fermionic phase space variables.
If we now introduce
the notation $x_1=q_1, x_2=p_1, x_3=q_2, x_4=p_2,..., x_{2N-1}=
q_N, x_{2N}=p_N$ for the operator phase space variables,
Eq.~(8) can be compactly rewritten as
$$\{ {\bf A}, {\bf B} \} = {\bf Tr} \sum_{r,s=1}^{2N}
\left( {\delta {\bf A} \over \delta x_r} \omega_{rs}
{\delta {\bf B} \over \delta x_s }  \right) ~~~,\eqno(9a)$$
and similarly, the operator Hamilton equations of Eq.~(5) can be
compactly rewritten as
$$\dot x_r=\sum_{s=1}^{2N} \omega_{rs} {\delta {\bf H} \over \delta x_s} ~~~.
\eqno(9b)$$
(Henceforth, we will not explicitly indicate the range of summation indices;
the index $r$ on $q_r,~p_r$ will be understood to have an upper summation
limit of $N$, while the index $r$ on $x_r$ will be understood to have an upper
summation limit of $2N$.)
If for convenience we order the bosonic variables
before the fermionic ones in the $2N$ dimensional phase space vector $x_r$,
then the matrix $\omega_{rs}$ which appears in Eqs.~(9a, b) is given by
$$\omega= {\rm diag}(\Omega_B,...,\Omega_B,\Omega_F,....,\Omega_F)~~~,
\eqno(10a)$$
with the $2 \times 2$ matrices $\Omega_{B,F}$ given by
$$\Omega_B=\left(\matrix{0&1\cr-1&0\cr}\right)~~,~~~
\Omega_F=- \left(\matrix{0&1\cr1&0\cr}\right)~~~,\eqno(10b)$$
and $\omega$ obeys
$$\eqalign{
&(\omega^2)_{rs} = -\epsilon_r \delta_{rs}~~,~~~\omega^4=1~~,~~~
\omega_{sr}=-\epsilon_r \omega_{rs}=-\epsilon_s \omega_{rs}~~~, \cr
&\sum_r \omega_{rs} \omega_{rt} = \sum_r \omega_{sr} \omega_{tr} =
 \delta_{st}~~~. \cr}\eqno(10c)$$
This concludes our review of generalized quantum dynamics; the reader
interested in further details, including the Lagrangian formulation and
some concrete examples of models constructed using the total trace formalism,
should consult Refs.~[1--4].
\vfill
\eject
\bigskip
\centerline{{\bf 2.~~Weyl Ordered Hamiltonians}}

Let us now consider a special class of operator Hamiltonians called
{\it Weyl ordered} Hamiltonians, in which the bosonic operators are all
totally symmetrized with respect to one another and to the fermionic
operators, and in which the fermionic operators are totally antisymmetrized
with respect to one another.  Clearly, the most general Weyl ordered
Hamiltonian which is a polynomial in the operator phase space variables
$\{x_r\}$ will be a sum of terms, which may be of different degrees,
each obtained by Weyl ordering a distinct monomial in the
phase space variables.  The contribution of all such monomials of degree $n$
may be simply represented by a generating function $G_n$ constructed as
follows.
Let $\sigma_r,~~r=1,...,2N$ be a set of parameters which are real numbers
when $\epsilon_r=1$ and which are real Grassmann numbers, which anticommute
with each other and with all of the fermionic phase space variables,
when $\epsilon_r=-1$.  Then if we form
$$G_n=g^n~~,~~~g=\sum_s \sigma_s x_s~~~, \eqno(11a)$$
the coefficient of each distinct monomial in the parameters $\sigma_r$ will
be a distinct Weyl ordered polynomial of degree $n$ in the phase space
variables $\{x_r\}$.
Corresponding to the operator
generating function $G_n$,  we define a total trace functional generating
function
$${\bf G}_n= {\bf Tr} G_n~~~,\eqno(11b)$$
where we specify the action of
${\bf Tr}$ on the Grassmann parameters in $G_n$ by requiring that
each Grassmann $\sigma_r$ anticommutes with $(-1)^F$.
The part of ${\bf G}_n$ which is even in the Grassmann parameters
is then a generating function for all nonvanishing
total trace functionals that correspond to the bosonic Weyl ordered monomials
generated by $G_n$.

Let us now compare the generalized quantum dynamics equations
of motion produced by
${\bf G}_n$ for general operators $\{ x_r \}$,
with the corresponding Heisenberg picture equations of motion
produced by $G_n$ when the phase space variables $\{ x_r \}$ are
assumed to obey the canonical algebra of complex quantum mechanics.  In
our compact phase space notation, this algebra takes the form
$$ \eqalign{
&x_r x_s-\epsilon_r x_s x_r = i \epsilon_r \omega_{rs}~~~, \cr
&[x_r, i]=0~~~, \cr
}\eqno(12)$$
where we adopt the convention that if only {\it one} of $x_r, x_s$ is bosonic,
it is taken to be the operator $x_r$; alternatively, we can rewrite the
first part of Eq.~(12) with no restrictions on the indices $r,s$ by including
a factor $\sigma_s$, giving
$$[x_r, \sigma_s x_s]= i \omega_{rs} \sigma_s~~~.
\eqno(13)$$

Applying the equations of motion of Eq.~(9b) with ${\bf G}_n$ playing
the role of the total trace Hamiltonian, we get
$$  \dot x_r = \sum_s \omega_{rs} {\delta {\bf G}_n \over \delta x_s}
= \sum_s \omega_{rs} n g^{n-1} \sigma_s~~~. \eqno(14)$$
On the other hand, from the canonical algebra of Eq.~(13) we find, for
both bosonic and fermionic $x_r$, that
$$[x_r,g]= i \sum_s \omega_{rs} \sigma_s~~~,\eqno(15a)$$
which in turn implies that
$$[x_r, G_n] = n g^{n-1} i \sum_s \omega_{rs} \sigma_s~~~. \eqno(15b)$$
But the Heisenberg picture equations of motion for the phase space variables,
taking $G_n$ as the operator Hamiltonian, are
$$\dot x_r=i [G_n,x_r]~~~, \eqno(15c)$$
which substituting Eq.~(15b) becomes
$$\dot x_r= \sum_s \omega_{rs} n g^{n-1} \sigma_s~~~, \eqno(16a)$$
in agreement with Eq.~(14).  We can now sum over all generating
function contributions $G_n$ weighted by $c$--number coefficients
to obtain a general Weyl ordered Hamiltonian $H$,
which has a corresponding total trace Hamiltonian ${\bf H}= {\bf Tr}H$,
which respectively generate the Heisenberg picture equation of motion
$$\dot x_r= i [H,x_r] \eqno(16b)$$
and the corresponding generalized quantum dynamics equation of motion of
Eq.~(9b).

Thus, for Weyl ordered Hamiltonians formed with $c$-number coefficients,
we conclude that the generalized quantum dynamics
equations of motion generated by {\bf H}
agree with the Heisenberg picture equations of motion generated by
$H$, on an
initial time slice on which the phase space variables are canonical,
and that on this time slice
$$[H, i]=0 ~~~. \eqno(16c)$$
But since Eq.~(16c) guarantees that
the Heisenberg picture equations of motion preserve
the canonical algebra on the next time slice, integrating forward in time
step by step then implies that generalized quantum
dynamics agrees with Heisenberg picture dynamics at all subsequent times,
and therefore defines a unitary dynamics.
We have given the argument here in a form which applies in complex
Hilbert space, where a $c$--number is a general complex number;
in quaternionic Hilbert space $i$ in the above equations is actually an
operator in the left quaternion algebra, denoted by $I$ in [1, 2], and the
most general $c$--number is a real number; while in (even dimensional) real
Hilbert space $i$ is represented by a $2 \times 2$ real matrix $i_2$,
which has the form of $-\Omega_B$ of Eq.~(10b), and the most general
$c$--number is again a real number.

When $H$ is not Weyl ordered, one can give explicit examples in which
the generalized quantum dynamics equations of motion do not
agree with those computed
from Heisenberg picture quantum mechanics.  With one pair of bosonic
phase space
variables $q,p$, for example, and the fifth degree Hamiltonian
$$\eqalign{
H=&\gamma_1 (p^3q^2+p^2q^2p+pq^2p^2+q^2p^3+qp^3q) \cr
      +&\gamma_2 (p^2qpq+pqpqp+qpqp^2+pqp^2q+qp^2qp) ~~~, \cr
      }\eqno(17a)$$
explicit calculation gives
$$[iH, q]- {\delta {\bf H} \over \delta p}= 2(\gamma_1-\gamma_2)~~~,
\eqno(17b) $$
which vanishes only in the Weyl ordered case $\gamma_1=\gamma_2$.  The
nonvanishing contribution to the right hand side of Eq.~(17b) can be
traced to the operator rearrangement
$$(q^2p^2-qp^2q)+(p^2q^2-qp^2q)=2iqp-2ipq=-2~~~,\eqno(17c)$$
which involves two successive applications of the canonical
commutator $[q,p]=i$.  Taking the trace of the left and right hand sides
of Eq.~(17c) clearly leads to a contradiction if cyclic invariance of the
trace is assumed.  This example serves as a warning that, even though
taking ${\bf Tr}$ of Eq.~(12) does not directly lead to an inconsistency
(because while
the cyclic property of the trace implies that
$$ {\bf Tr} (x_r x_s -\epsilon_r x_s x_r)
= {\bf Tr} [x_r x_s - (\epsilon_r)^2 x_r x_s]=0~~~, \eqno(17d)$$
the inclusion of the real part ${\rm Re}$
in the definition of Eq.~(1) makes ${\bf Tr} i=0$), the canonical algebra is
in general inconsistent with the cylic trace property.  {\it Hence we cannot
simply impose the canonical algebra by fiat as an initial condition in
generalized quantum dynamics!}  But we shall see that an effective canonical
algebra can arise as an {\it emergent property} of ensemble averages
in the statistical mechanics of generalized quantum dynamics.

To complete the analysis of this section, we note that Weyl ordering is
not a necessary condition for the two forms of
dynamics to agree, as can be seen by considering the fourth degree
Hamiltonian
$$H=\gamma_1(p^2q^2+q^2p^2)+\gamma_2pq^2p +\gamma_3qp^2q
+\gamma_4(pqpq+qpqp)~~~, \eqno(17e)$$
for which one finds that the two forms of dynamics coincide for arbitrary
values of the coefficients $\gamma_{1,...,4}$ which multiply distinct
self--adjoint combinations of operators.  This example, and analogs with
more than one degree of freedom, are relevant for the
behavior of the operator gauge invariant extensions of standard gauge
theories formulated in Refs. [1, 2], which we will study in detail elsewhere.

\bigskip
\centerline{{\bf 3.~~The Conserved Operator $ \tilde C$}}

Let us now make a further application of the generating function ${\bf G}_n$
of Eq.~(11b).  Taking the operator derivative with respect to $x_s$ and using
the fact that $\sigma_s$ commutes with $g$, we have
$$\eqalign{
{\delta {\bf G}_n \over \delta x_s}=&n g^{n-1} \sigma_s  \cr
=&\sigma_s n g^{n-1}~~.  \cr}
\eqno(18)$$
Multiplying the first equality in Eq.~(18) by $x_s$ from the right and
summing over $s$, we get
$$\sum_s {\delta {\bf G_n} \over \delta x_s} x_s= n g^{n-1} \sum_s \sigma_s
x_s= n g^n~~~, \eqno(19a)$$
while multiplying the second equality in Eq.~(18) by $\epsilon_s x_s$ from
the left and summing over $s$, we get
$$\sum_s \epsilon_s x_s {\delta {\bf G_n} \over \delta x_s} =
\left(\sum_s \epsilon_s x_s \sigma_s\right) n g^{n-1}=
\left( \sum_s \sigma_s x_s \right) n g^{n-1} = n g^n~~~. \eqno(19b)$$
Since the right--hand sides of Eqs.~(19a, b) are the same, subtracting
them gives
$$ \sum_s \left( {\delta {\bf G}_n \over \delta x_s} x_s
-\epsilon_s x_s {\delta {\bf G}_n \over \delta x_s} \right) =0~~~,
\eqno(19c)$$
which when summed with $c$--number coefficients over all monomial
contributions to the Weyl ordered total trace Hamiltonian {\bf H} yields
the important identity
$$ \sum_s \left( {\delta {\bf H} \over \delta x_s} x_s
-\epsilon_s x_s {\delta {\bf H} \over \delta x_s} \right) =0~~~.
\eqno(20)$$

The identity of Eq.~(20) is in fact more general than is suggested by the
preceding derivation, and holds even if ${\bf H}= {\bf Tr} H$ is
{\it not} Weyl ordered, provided only that $H$ is constructed from
monomials formed from the $\{ x_r \}$ using only coefficients that commute
with all bosonic operators in Hilbert space (that is, as discussed above,
real coefficients in quaternionic and
real Hilbert space and complex coefficients in complex Hilbert space; in
addition to these $c$--numbers, the coefficients can also depend on the
grading operator $(-1)^F$.)
To see this, let us consider two distinct variations of {\bf H} which
we label $\delta_1$ and $\delta_2$, defined respectively by
$$\eqalign{
\delta_1 x_r=&x_r \delta \Lambda~~~, \cr
\delta_2 x_r=&(\delta \Lambda) x_r~~~, \cr }
\eqno(21)$$
with $\delta \Lambda$ an arbitrary infinitesimal
anti--self--adjoint bosonic operator variation.
As long as the only noncommutativity in $H$ arises from the phase
space variables $\{ x_r \}$, which is the case when only $c$--number
coefficients are employed in constructing $H$, cyclic invariance
of {\bf Tr} implies that the two variations give the same result when
applied to {\bf H}, since the term ${\cal O}_L \delta \Lambda {\cal O}_R$
arising from $\delta_1$ acting on the right--most operator in ${\cal O}_L$
is identical to the term arising from $\delta_2$ acting on the left--most
operator in ${\cal O}_R$.  Applying Eq.~(4) to Eq.~(21) gives
$$\eqalign{
0=&{\bf Tr} \sum_s {\delta {\bf H} \over \delta x_s} [x_s \delta \Lambda
-(\delta \Lambda) x_s]  \cr
=&{\bf Tr} \sum_s \left[{\delta {\bf H} \over \delta x_s} x_s
-\epsilon_s x_s {\delta {\bf H} \over \delta x_s} \right] \delta \Lambda~~~,
\cr}
\eqno(22a)$$
which since $\delta \Lambda$ is an arbitrary anti--self--adjoint
bosonic operator implies the anti--self--adjoint operator relation
$$0=\sum_s \left[{\delta {\bf H} \over \delta x_s} x_s
-\epsilon_s x_s {\delta {\bf H} \over \delta x_s} \right] ~~~,
\eqno(22b)$$
giving the same identity as was obtained in the Weyl ordered case in
Eq.~(20).

We shall now rewrite the identity of Eq.~(22b) in an alternative useful
form.  Let us define the operator $\tilde C$ by
$$\eqalign{
\tilde C=&\sum_{r,s} x_r \omega_{rs} x_s  \cr
=&\sum_{r,B} [q_r, p_r] - \sum_{r,F} \{q_r, p_r\}~~~; \cr }
\eqno(23)$$
that is, $\tilde C$ is the difference between the sums of bosonic
commutators and fermionic anticommutators.  (The
tilde is a reminder that $\tilde C$ is anti--self--adjoint, a point that
will be discussed in detail shortly.)  Differentiating the first line of
Eq.~(23) with respect to time, and using the Hamilton equations of Eq.~(9b)
and the properties of $\omega_{rs}$ summarized in Eq.~(10c), we find
$$\eqalign{
\dot{\tilde C}=&\sum_{r,s}( \dot x_r \omega_{rs} x_s +
x_r \omega_{rs} \dot x_s )  \cr
=&\sum_{r,s,t} \left( {\delta {\bf H} \over \delta x_t} \omega_{rt}
\omega_{rs} x_s + x_r \omega_{rs}
\omega_{st} {\delta {\bf H} \over \delta x_t}  \right) \cr
=&\sum_r \left( {\delta {\bf H} \over \delta x_r} x_r
-\epsilon_r x_r {\delta {\bf H} \over \delta x_r} \right) \cr
=&0~~~, \cr}
\eqno(24)$$
where in the final equality we have used the identity of Eq.~(22b).
Thus the operator $\tilde C$ is a constant of the motion for generalized
quantum dynamics, as long as the total trace Hamiltonian {\bf H} is
constructed from the operator phase space variables using only
coefficients that commute with all bosonic phase space operators.

We will exploit the conservation of $\tilde C$ in the following two sections,
but pause to make two remarks.\hfill \break
(1)  We first comment on the adjointness properties of $\tilde C$.  Taking
the bosonic coordinates $\{ q_r \}$ and conjugate momenta
$\{ p_r \}$ to be self--adjoint operators, the bosonic commutator terms
in Eq.~(23) defining $\tilde C$ are evidently anti--self--adjoint.  (In
Refs.~[1, 2] the possibility of anti--self--adjoint bosonic $\{q_r\}$ and
$\{p_r\}$ was considered, and this adjointness assignment
also leads to an anti--self--adjoint $\tilde C$; however, because of the
complex quantum mechanical structure of the final results of this paper
we expect the anti--self--adjoint case to resemble the self--adjoint case,
and do not consider it further.)
In complex Hilbert space, where $i$ is a $c$--number,
the usual fermionic Lagrangians lead to the
identifications (see Sec. 13.6 of Ref. [2])
$q_r = \psi_r,~~~p_r=i \psi_r^{\dagger} $, with
$\psi_r$ a fermionic operator which is neither self--adjoint nor
anti--self--adjoint. This
gives $q_r^{\dagger} = -i p_r,~~~p_r^{\dagger}=-i q_r$,
as a consequence of which the fermionic anticommutator terms in Eq.~(23) are
also anti--self--adjoint.  An analogous construction
also applies in quaternionic
Hilbert space, with the left algebra operator $I$ playing the role of $i$.
One must now pay attention to factor ordering and use of a manifestly
self--adjoint Lagrangian, giving for each fermionic degree of freedom a pair
of phase space operators $q_{r1}=\psi_r$, $p_{r1}={1 \over 2}
\psi_r^{\dagger}I$,
$q_{r2}={1 \over 2} I \psi_r$, and $p_{r2}=\psi_r^{\dagger}$,
so that $p_{r2}^{\dagger}=q_{r1}~,~~q_{r2}^{\dagger}=-p_{r1}$, which makes
$\{q_{r1}, p_{r1} \} + \{q_{r2}, p_{r2} \}$ anti--self--adjoint.
In quaternionic Hilbert space, the only way
to construct a total trace Lagrangian for fermions without
using an explicit imaginary unit is to introduce the fermions
in pairs, with the real matrix $-\Omega_B$ introduced above playing the
role of the imaginary unit (see Sec. 13.7 of Ref.~[2]).  For each $r$
one then has $q_{r1}, p_{r1}, q_{r2}, p_{r2}$, with adjointness properties
assigned according to
$p_{r2}^{\dagger}=\pm q_{r1}~,~~q_{r2}^{\dagger}=\mp p_{r1}$
(but now with no relation between $p_{r1}$ and $q_{r1}$),
which again makes the two--term sum $\{q_{r1},p_{r1}\} + \{q_{r2},p_{r2} \}$
anti--self--adjoint.  This fermionic construction also applies to real
Hilbert space.  \hfill \break
(2)  Not surprisingly, the conservation of $\tilde C$ can be given a
Noether's theorem formulation in terms of the total trace Lagrangian
${\bf L}={\bf Tr} L[\{q_r\},\{\dot q_r\}]$ which
corresponds to {\bf H}.  We discuss here the simplest case, in which
$L$ is constructed from its
operator arguments using only $c$--number coefficients, and involves
no constraints.  Cyclic invariance
of {\bf Tr} then implies that {\bf L} is invariant under the operator
variations $\delta q_r=\delta_2 q_r - \delta_1 q_r =
[\delta \Lambda,q_r]~,~~~r=1,...,N$ for a
time--independent bosonic variation $\delta \Lambda$.  The generalization
of Noether's theorem to total trace Lagrangians given in Sec.~13.5 of
Ref.~[2] then implies that there is a conserved charge $Q_{\Lambda}$
obeying $\dot Q_{\Lambda}=0$ and given by
$$\eqalign{
Q_{\Lambda}=&{\delta {\bf L} \over \delta {\dot \Lambda}} \cr
=&\sum_{r,B} \left[q_r, {\delta {\bf L} \over \delta \dot q_r} \right]
-\sum_{r,F}  \left\{q_r, {\delta {\bf L} \over \delta \dot q_r} \right\}~~~,
\cr
}
\eqno(25)$$
which on substituting $p_r = \delta {\bf L} /{ \delta \dot q_r}$ becomes
identical to $\tilde C$.  If the label $r$ is a composite label comprising
a spatial coordinate $\vec x$ as well as a discrete field index $r$,
then the
Noether's theorem argument implies that there is a current $J^{\mu}$
which obeys $\partial_{\mu} J^{\mu}=0$ and is given by
$$J^{\mu}=\sum_{r,B}
\left[q_r, {\delta {\bf L} \over \delta \partial_{\mu} q_r} \right]
-\sum_{r,F}
\left\{q_r, {\delta {\bf L} \over \delta \partial_{\mu} q_r} \right\}~~~,
\eqno(26)$$
with $\tilde C= \int d^3x ~J^0$ the associated charge.
\vfill\eject
\centerline{{\bf 4.~~Canonical Transformations }}

We turn now to an analysis of the structure of canonical transformations
and symmetry transformations in generalized quantum dynamics.  Generalizing
from the structure [5] of infinitesimal canonical transformations in
classical mechanics, an infinitesimal canonical transformation
in generalized quantum dynamics is defined by
$$x_r^{\prime}-x_r \equiv \delta x_r = \sum_s \omega_{rs}
{ \delta {\bf G} \over \delta x_s} ~~~. \eqno(27)$$
Here ${\bf G} = {\bf Tr} G$, with G self--adjoint, is a total
trace functional constructed from the phase space operators $\{ x_r \}$
using arbitrary  coefficients, which can be fixed operators as well as
$c$--numbers.  When Eq.~(27) is restricted by requiring that the
coefficients used to form $G$ are composed of
either $c$--numbers, Grassmann $c$--numbers,
or the grading operator $(-1)^F$, the transformation will be termed an
{\it intrinsic} canonical transformation, and when the further condition of
a Weyl ordered $G$ is imposed, the transformation will be termed a
{\it Weyl ordered intrinsic} canonical transformation.

Letting ${\bf A} \equiv {\bf A}[\{ x_r \}]$
be an arbitrary total trace functional, we find immediately that to first
order under a canonical transformation,
$$\eqalign{
{\bf A}^{\prime} \equiv& {\bf A} [\{ x_r^{\prime} \}] \cr
=&{\bf A} +{\bf Tr} \sum_r {\delta {\bf A} \over \delta x_r}
\delta x_r \cr
=&{\bf A} +{\bf Tr} \sum_{r,s} {\delta {\bf A} \over \delta x_r}
\omega_{rs}  { \delta {\bf G} \over \delta x_s} \cr
=&{\bf A} +\{ {\bf A}, {\bf G} \}~~~, \cr
} \eqno(28a)$$
that is,
$$\delta {\bf A} \equiv
{\bf A}^{\prime}-{\bf A}= \{ {\bf A}, {\bf G} \}~~~.\eqno(28b)$$
Comparing Eq.~(27) with Eq.~(9b), we see that when {\bf G}
is taken as ${\bf H}dt$, with {\bf H}
the total trace Hamiltonian and $dt$ an infinitesimal time step,
then $\delta x_r =\dot x_r dt$ gives the small change in $x_r$
resulting from
the dynamics of the system over that time step.
 From Eq.~(28a), we see that when ${\bf A}={\bf Tr}A$ with $A$ a
Weyl ordered polynomial in the arguments $\{ x_r \}$, the canonically
transformed total trace functional ${\bf A}+ \{ {\bf A}, {\bf G} \}$
obtained by applying a
canonical transformation is again a
Weyl ordered polynomial in the new arguments $\{ x_r^{\prime} \}$.
However, since we shall see in Appendix F (where we discuss further details
of canonical transformations) that the class of Weyl ordered
total trace functionals is {\it not} closed under the generalized Poisson
bracket operation, the transformed functional ${\bf A} ^{\prime}$ is not
in general a Weyl ordered polynomial in the original arguments
$\{ x_r  \}$.

In Sec.~2 we saw that for Weyl ordered Hamiltonians, there is a close
relationship between the time evolution under generalized quantum dynamics
and the corresponding Heisenberg dynamics generated when the operator
variables are assumed to obey canonical commutators.
Generalizing the
calculation of Eqs.~(14--16b) to the case when
{\bf H} is replaced by any Weyl ordered intrinsic canonical generator
{\bf G} [the use now of Grassmann $c$--number coefficients in
adding monomials causes no problems since $g$ in Eq.~(11a) is bosonic],
we see that in this case Eq.~(27) can be
represented over the canonical algebra
by a commutator as well as by an operator derivative,
$$\delta x_r = \sum_s \omega_{rs}
{ \delta {\bf G} \over \delta x_s}
=i[G,x_r]   ~~~,\eqno(29)$$
with the first equality in Eq.~(29) holding for arbitrary operator arguments,
and the second equality holding only over the canonical algebra.

In the remainder of this section we establish two important invariances under
canonical transformations in generalized quantum dynamics.
We consider
first the change in the conserved operator $\tilde C$ under an intrinsic
canonical transformation, giving by use of Eq.~(27)
$$\eqalign{
\delta \tilde C=& \sum_{r,s}[ \delta x_r \omega_{rs} x_s+
x_r \omega_{rs} \delta x_s]  \cr
=&\sum_{r,s,t} \left(  {\delta {\bf G} \over \delta x_t} \omega_{rt}
\omega_{rs} x_s + x_r \omega_{rs}
\omega_{st} {\delta {\bf G} \over \delta x_t}  \right) \cr
=&\sum_r \left( {\delta {\bf G} \over \delta x_r} x_r
-\epsilon_r x_r {\delta {\bf G} \over \delta x_r} \right)~~~, \cr
}\eqno(30)$$
where we have again used the identities of Eq.~(10c).
We recognize that the right hand side of Eq.~(30) has the same structure as
was encountered in Sec.~3, with {\bf G} now playing the role of {\bf H}.
Therefore by the same cyclic invariance argument as was used previously in
Eqs.~(21--22), we conclude that as long as only $c$--numbers,
Grassmann $c$--numbers, or the grading operator $(-1)^F$, all of which
commute with an
arbitrary bosonic $\delta \Lambda$, are
used in constructing {\bf G}, the right hand side of Eq.~(30) vanishes
and $\tilde C$ is intrinsic canonical invariant.

The second invariance concerns the phase space measure for the phase space
operators $\{ x_r \}$.  Let us introduce a complete set of states
$\{ |n \rangle \}$ in the underlying Hilbert space, so that the phase
space operators are completely characterized by their matrix elements
$\langle m| x_r |n \rangle \equiv (x_r)_{mn}$, which have the
following form in real, complex, and quaternionic Hilbert space:
$$\eqalign{
&{\rm In~ real~ Hilbert~ space:~~}
(x_r)_{mn} =(x_r)_{mn}^0 \cr
&{\rm In~ complex~ Hilbert~ space:~~}
(x_r)_{mn}=(x_r)_{mn}^0+i(x_r)_{mn}^1 \cr
&{\rm In~ quaternionic~ Hilbert~ space:~~}
(x_r)_{mn}=(x_r)_{mn}^0+i(x_r)_{mn}^1+j(x_r)_{mn}^2+k(x_r)_{mn}^3
{}~~~,\cr
} \eqno(31)$$
with $(x_r)_{mn}^A~,~~A=0,1,2,3$ real numbers.  (Note
that this is true for fermionic as well as bosonic operators; the matrix
elements of fermionic operators are still real numbers, {\it not} real
Grassmann numbers!  Grassmann numbers are employed only as auxilliary
quantities in forming Weyl ordered products and in grade--changing
transformations.)  If for the moment we ignore adjointness restrictions,
the phase space measure is defined by
$$\eqalign{
d\mu=&\prod_A d \mu^A~~~, \cr
d\mu^A \equiv& \prod_{r,m,n} d(x_r)_{mn}^A~~~;\cr
}\eqno(32)$$
when adjointness restrictions are taken into account, certain factors in
Eq.~(32) become redundant and are omitted.  Our strategy is to first
ignore adjointness restrictions and to prove the canonical invariance of
each individual factor $d\mu^A$ in the first line of Eq.~(32); in
Appendix A we
describe how the adjointness restrictions modify Eq.~(32) and show
that the proof remains valid when these modifications are taken into account.

Under the canonical transformation of Eq.~(27), the matrix elements
of the new variables $x_r^{\prime}$ are related to those of the original
variables $x_r$  by
$$(x_r^{\prime})_{mn}^A=(x_r)_{mn}^A+ \sum_s  \omega_{rs}
\left( {\delta {\bf G} \over \delta x_s} \right)_{mn}^A~~~. \eqno(33)$$
Inserting a complete set of intermediate states into the fundamental
definition
$$\delta {\bf G} = {\bf Tr} \sum_s {\delta {\bf G} \over \delta x_s}
\delta x_s~~~,  \eqno(34)$$
we get
$$\delta {\bf G} =  \sum_{s,m,n,A} \epsilon_m
\epsilon^A \left( {\delta {\bf G} \over \delta x_s} \right)_{mn}^A
(\delta x_s)_{nm}^A~~~,\eqno(35)$$
where $\epsilon_m=1(-1)$ according to whether the state $|m \rangle$ is
bosonic (fermionic), and where $\epsilon^0=1$ and $\epsilon^A=-1,~~~A=1,2,3$.
(In Refs.~[1, 2] the
factor of $\epsilon^A$ was inadvertently omitted, but this does not affect
the alternative proof of the Jacobi identity given there.)  Thus, we see that
$$\left( {\delta {\bf G} \over \delta x_s} \right)_{mn}^A =
\epsilon_m \epsilon^A {\partial {\bf G} \over \partial (x_s)_{nm}^A }
{}~~~,\eqno(36)$$
allowing us to rewrite Eq.~(33) in terms of ordinary partial derivatives of
the total trace functional {\bf G},
$$(x_r^{\prime})_{mn}^A=(x_r)_{mn}^A+ \sum_s  \omega_{rs}
\epsilon_m \epsilon^A {\partial {\bf G} \over \partial (x_s)_{nm}^A }
{}~~~. \eqno(37)$$
Differentiating Eq.~(37) with respect to $(x_{r^{\prime}})_{m^{\prime}
n^{\prime}}^A$, we get for the transformation matrix
$${\partial (x_r^{\prime})_{mn}^A  \over \partial
(x_{r^{\prime}})_{m^{\prime} n^{\prime}}^A  }=
\delta_{r r^{\prime}} \delta_{m m^{\prime}} \delta_{n n^{\prime}}
+ \sum_s \omega_{rs} \epsilon_m \epsilon^A
{ \partial^2 {\bf G} \over \partial (x_s)_{nm}^A
\partial (x_{r^{\prime}})_{m^{\prime} n^{\prime}}^A  }  ~~~. \eqno(38)$$
Since for an infinitesimal matrix $\delta X$ we have $\det(1+\delta X)
\approx 1+{\rm Tr} \delta X$, we learn from Eq.~(38) that the Jacobian of
the transformation is
$$\eqalign{
J=&1+\Sigma~~~, \cr
\Sigma=&\sum_{r,s,m,n}
\omega_{rs} \epsilon_m \epsilon^A
{ \partial^2 {\bf G} \over \partial (x_s)_{nm}^A \partial (x_r)_{mn}^A  } \cr
}~~~. \eqno(39)$$
Interchanging in $\Sigma$ the summation indices $r$ and $s$,  and
also interchanging
the summation indices $m$ and $n$, we get
$$
\Sigma=\sum_{r,s,m,n}
\omega_{sr} \epsilon_n \epsilon^A
{ \partial^2 {\bf G} \over \partial (x_r)_{mn}^A \partial (x_s)_{nm}^A  }
{}~~~, \eqno(40)$$
but now using $\omega_{sr} = -\epsilon_r \omega_{rs}$ together with the
relation $\epsilon_r = \epsilon_m \epsilon_n$ [which expresses the fact that
bosonic (fermionic) operators can only connect states of like (unlike)
fermion number], we obtain
$$
\Sigma=-\sum_{r,s,m,n}
\omega_{rs} \epsilon_m \epsilon^A
{ \partial^2 {\bf G} \over \partial (x_r)_{mn}^A \partial (x_s)_{nm}^A  }
{}~~~. \eqno(41)$$
But since the order of second partial derivatives with respect to real
matrix elements is immaterial, this is just the statement $\Sigma=-\Sigma$;
hence $\Sigma$
vanishes and the Jacobian of the transformation is unity.  Although we have
ignored adjointness restrictions in this argument, as shown in Appendix A the
conclusion is unaltered when these are taken into account.

To summarize, we have shown that the operator phase space integration
measure $d \mu$ is invariant under canonical transformations.
An important corollary of this result follows when {\bf G}
is taken as the generator $dt {\bf H}$ of an infinitesimal time translation,
since we then learn that $d \mu$ is invariant under the dynamical evolution
of the system, giving a generalized quantum dynamics analog of Liouville's
theorem of classical mechanics.  Since
no restrictions on the form of the generator {\bf G} were needed in the
above argument for the invariance of $d \mu$, the argument applies even when
{\bf G} is formed from the operator phase space variables
using {\it operator} coefficients.  Thus,
the integration measure $d \mu$ is invariant under a
unitary transformation on the basis of states in Hilbert space, the effect of
which on the variables $\{ x_r \}$ can be represented by Eq.~(27) with
${\bf G}=-{\bf Tr} \sum_r [\tilde G,p_r]q_r$, with $\tilde G$ a fixed bosonic
anti--self--adjoint operator.  This transformation, however, is
not an intrinsic
canonical transformation and is only a covariance, rather than an
invariance, of the operator $\tilde C$.
\bigskip
\centerline{{\bf 5.~~Equilibrium Ensemble of Operator Initial Values}}

The operator equations of motion of generalized quantum dynamics determine
the time evolution of the operator coordinates and momenta at all times,
given their values on an initial time slice.  However, these initial
values are themselves not determined.  We shall now make
the assumption that for a large enough system, the {\it statistical
distribution} of initial values can be treated by the methods of statistical
mechanics.  Specifically, we shall assume that the {\it a priori} distribution
of inital values is uniform over the operator phase space, so that the
equilibrium distribution is determined solely by maximizing the combinatoric
probability subject to the
constraints imposed by the generic conservation laws.  Liouville's
theorem implies that if the assumption of a uniform {\it a priori}
probability distribution is made at one time, then it
is valid at all later times, assuring the consistency of the concept of an
equilibrium ensemble.  We do not propose to address the question of how the
randomness in the initial value distribution arises:  It could come from a
random initial condition, an ordered initial condition
followed by evolution under an ergodic dynamics, or some combination of
the two.

More specifically, let $d\mu = d\mu[\{x_r\}]$ denote the operator phase
space measure discussed in detail in the preceding section.  In what follows
we shall not need the specific form of this measure, but only the properties
that it obeys Liouville's theorem, and that the
measure is invariant under infinitesimal operator shifts $ \delta x_r $,
that is
$$d\mu[\{x_r+ \delta x_r\}]= d\mu[\{x_r\}]~~~.\eqno(42)$$
(This property will not be used until Sec.~6, where we discuss the
equipartition or Ward identities.)
For a system in statistical equilibrium, there is an equilibrium
distribution of operator initial values $\rho[\{x_r\}]$, such that
$$ d P=d\mu[\{x_r\}]   \rho[\{x_r\}]  ~~~\eqno(43a)$$
is the infinitesimal probability of finding the system in the
operator phase space volume element $d \mu$, with the total probability
equal to unity,
$$1=\int  d P=\int d\mu[\{x_r\}]   \rho[\{x_r\}]  ~~~.\eqno(43b)$$
The first task in a statistical
mechanical analysis is to determine the equilibrium distribution $\rho$.

Since equilibrium implies that $\dot \rho=0$, the equilibrium distribution
can only depend on conserved operators and total trace functionals.
In the generic case for a Lorentz invariant system, the only conserved
operator is $\tilde C$ and the only conserved total trace functionals are
the total trace Hamiltonian ${\bf H}={\bf p}^0$, the total trace three
momentum ${\bf \vec p}$, and the total trace angular momentum ${\bf \vec J}$.
However, because the graded trace functionals are all
indefinite in sign, standard statistical methods lead to a divergent
partition function in the generic case.  We shall therefore restrict our
discussion to total trace Hamiltonians ${\bf H} ={\bf Tr} H$ for which
the generalized quantum dynamics equations of motion imply conservation of
the ungraded trace functional ${\bf \hat H }\equiv {\bf \hat {Tr}} H \equiv
{\rm Re} {\rm Tr} H$
as well as conservation of ${\bf H}$; we shall see
in Sec.~7 below that a large class of models has this property.  These
models are characterized (see Appendices C and G) by having an additional
conserved
operator $\tilde F$, which when restricted to the canonical algebra
corresponds to the operator for the conserved fermion number $F$.
In addition to the conservation of ${\bf \hat H}$, we shall assume
the more restrictive condition
that ${\bf \hat H}$ is bounded from below; conditions
for achieving this are also discussed in Sec.~7.  We shall also assume
henceforth an ensemble which is translation invariant,
rotation invariant, and Lorentz invariant.  Since $\tilde C$ is invariant
under intrinsic
canonical transformations, it is Lorentz invariant, and
so the equilibrium distribution
depends on $\tilde C$; similarly, since $\tilde F$ is
obtained from $\tilde C$ by a reordering of fermion factors it is
Lorentz invariant as well, and so the equilibrium
distribution also depends on $\tilde F$.
In the ensemble rest frame the equilibrium distribution
can also depend on ${\bf \hat H}$ and ${\bf H}$ (since these are the ungraded
and
graded mass functionals in the rest frame),
giving the general equilibrium  distribution
$$\rho=\rho(\tilde C,\tilde F, {\bf \hat H}, {\bf H})~~~.  \eqno(44a)$$
The analysis of the implications of this general equilibrium distribution
entails considerable algebraic complexity, the full details of which will be
published elsewhere.  However, as summarized in Appendices C and G,
the results are all qualitatively similar to those obtained from the
simpler equilibrium distribution
$$\rho=\rho(\tilde C,{\bf \hat H}, {\bf H})~~~,  \eqno(44b)$$
in which the $\tilde F$ dependence is dropped.  We shall focus on this
simplified case in the exposition that follows.

In addition to its dependence on the dynamical variables,
$\rho$ can also depend on constant parameter values, with
the functional form of $\rho$ and the values of the parameters together
defining a statistical ensemble.  Including an
anti--self--adjoint operator parameter $\tilde \lambda$
(which corresponds to the structure of $\tilde C$) and  real number
parameters $\hat {\tau}$ and $\tau$, which correspond to the structure of
${\bf {\hat H}}$ and ${\bf H}$, the
general form of the equilibrium ensemble corresponding to
Eq.~(44b) is
$$\rho=\rho(\tilde C, \tilde \lambda;{\bf{\hat H}},\hat {\tau};
{\bf H}, \tau)
{}~~~.\eqno(44c)$$
In the canonical ensemble, we shall see that the dependence on $\tilde C$ and
$\tilde \lambda$ is only through the single real number ${\bf Tr} \tilde
\lambda
\tilde C$, and so specializing to this case, Eq.~(44c) becomes
$$\rho=\rho({\bf Tr} \tilde \lambda \tilde C;{\bf{\hat H}},\hat{\tau};
{\bf H}, \tau)
{}~~~.\eqno(44d)$$

We shall now show that some significant consequences follow from the
general form of Eq.~(44d), together with the fact that the real
function $\rho$ on the right hand side of Eq.~(44d) is
constructed from its real number arguments using
only real number coefficients, and the assumption that ${\bf \hat H}$ and
${\bf H}$ are constructed from the operators $\{ x_r \}$
using only $c$--number coefficients and the grading operator $(-1)^F$.
For a general operator ${\cal O}$, let us define the ensemble average
$\langle {\cal O} \rangle_{AV}$ by
$$\langle {\cal O} \rangle_{AV}= {\int d\mu \rho {\cal O}  \over \int d\mu
\rho}~~~.\eqno(45a)$$
Then when ${\cal O}$ is constructed from the $\{ x_r \}$
using only $c$--numbers and $(-1)^F$ as coefficients,
the ensemble average $\langle  {\cal O} \rangle_{AV}$ must have the form
$$\langle {\cal O} \rangle_{AV}=F_{\cal O}(\tilde \lambda, (-1)^F)~~~,
\eqno(45b)$$
with the function $F_{\cal O}$ constructed from its arguments using
only $c$--number coefficients (in which we include the $\hat \tau$ and $\tau$
dependence).
This further implies that both the grading operator $(-1)^F$ and the ensemble
parameter $\tilde \lambda$ commute with $\langle {\cal O} \rangle_{AV}$,
$$[\tilde \lambda, \langle {\cal O} \rangle_{AV}] =
[(-1)^F, \langle {\cal O} \rangle_{AV}]=0~~~.\eqno(45c)$$

Let us now exploit the fact that the anti--self--adjoint operator
$\tilde \lambda$ can always be diagonalized (or, in the case of an even
dimensional real Hilbert space, reduced to $2 \times 2$
diagonal blocks), by a unitary transformation on
the basis of states in Hilbert space, which we have seen is
also an invariance of the
integration measure $d \mu$.  The functional relationship between
$\tilde \lambda$ and $\langle \tilde C \rangle_{AV}$ then implies that
$\langle \tilde C \rangle _{AV}$ is diagonal (or block diagonal)
in this basis as well.  As described more fully in Appendix B, this
brings $\langle \tilde C \rangle_{AV}$ into the following canonical form
in real (when even dimensional), complex, and quaternionic Hilbert space,
$$\eqalign{
&\langle \tilde C\rangle_{AV}=i_{eff} D~,~~~  i_{eff}=-i_{eff}^{\dagger}~,~~~
i_{eff}^2=-1~,~~~\cr
&[i_{eff},D]=0~,~~~ D~~ {\rm real~diagonal~and~nonnegative}~.~~~\cr
}\eqno(46a)$$
Although the case of general $D$ is interesting, we shall restrict
ourselves in this paper to the special case in which $D$ is a
real constant times the unit operator; in other words, we are assuming that
the ensemble does not favor any state in Hilbert space over any other.
Benefiting from some prescience, we denote this real constant by  $\hbar$,
and so have
$$\eqalign{
\langle \tilde C \rangle_{AV}=&i_{eff} \hbar~,~~~\cr
\{i_{eff}, \langle \tilde C \rangle_{AV} \}=&-2 \hbar~.~~~\cr
}\eqno(46b)$$

We turn now to the calculation of the functional form of $\rho$ in the
canonical ensemble, which is the ensemble relevant for describing the
behavior of a large system which is a subsystem of a still larger system.
The form of $\rho$ is determined [6, 7]
by minimizing the negative of the entropy,
$$-S=\int d\mu \rho \log \rho ~,~~~\eqno(47a)$$
subject to the constraints
$$\eqalign{
\int d\mu \rho=&1~,~~~\cr
\int d\mu \rho \tilde C =&\langle \tilde C \rangle_{AV}~,~~~\cr
\int d\mu \rho {\bf \hat {Tr}} H =&\langle {\bf \hat{Tr}} H
\rangle_{AV}~,~~~\cr
\int d\mu \rho {\bf Tr} H=&\langle {\bf Tr}H \rangle_{AV}~.~~~\cr
}\eqno(47b)$$
The standard procedure is to impose the constraints with Lagrange
multipliers by writing
$${\cal F}=\int d\mu\rho \log \rho+\theta \int d\mu \rho + \int d\mu \rho
{\bf Tr} \tilde \lambda \tilde C +\hat{\tau} \int d\mu \rho {\bf \hat{Tr}}H
+\tau \int d\mu \rho {\bf Tr} H~~~,  \eqno(48a)$$
and minimizing ${\cal F}$, treating all variations of $\rho$ as independent.
(Note that it makes no difference whether the constraint for $\tilde C$ is
introduced through a graded or an ungraded trace; the difference is a factor
of $(-1)^F$ which can be absorbed into the definition of $\tilde \lambda$.)
In order for ${\cal F}$ to have a minimum, it must be bounded
below; we shall assume that this is the case for sufficiently large
$\hat{\tau}$
(at a minimum, one needs $\hat{\tau} \ge |\tau|$, so that the coefficients
of the trace of $H$ over the bosonic and fermionic subspaces, proportional
respectively to $\hat \tau +\tau$ and $\hat \tau -\tau$, are both
positive). Varying Eq.~(48a) with respect to $\rho$ then gives
$$\rho=\exp(-1-\theta -{\bf Tr}\tilde \lambda \tilde C-\hat{\tau}
{\bf \hat{Tr}}H
-\tau {\bf Tr}H)~,~~~
\eqno(48b)$$
which on imposing the condition that $\rho$ be normalized to unity gives
finally
$$\eqalign{
\rho=&Z^{-1}\exp(-{\bf Tr}\tilde \lambda\tilde C -\hat{\tau} {\bf \hat{Tr}}H
-\tau {\bf Tr}H)~,~~~\cr
Z=&\int d\mu\exp(-{\bf Tr}\tilde \lambda\tilde C -\hat{\tau} {\bf \hat{Tr}}H
-\tau {\bf Tr}H) ~.~~~\cr
}\eqno(48c)$$

 From Eq.~(48c) we can easily derive some elementary statistical properties
of the equilibrium ensemble.  For the entropy $S$, we find
$$S=-\int d\mu \rho \log \rho = \log Z + {\bf Tr} \tilde \lambda
\langle \tilde C \rangle_{AV} + \hat{\tau} \langle {\bf \hat{Tr}} H
\rangle_{AV}
+ \tau \langle {\bf Tr} H \rangle_{AV}~.~~~
\eqno(49a)$$
Since the ensemble averages which appear in Eq.~(49a) are given by
$$\eqalign{
\langle \tilde C \rangle_{AV}=&-{\delta \log Z \over \delta \tilde \lambda} ~,
{}~~~\cr
\langle {\bf\hat{Tr}} H \rangle_{AV}=&-{\partial \log Z \over \partial
\hat{\tau}}~,~~~\cr
\langle {\bf Tr} H \rangle_{AV}=&-{\partial \log Z \over \partial
\tau}~,~~~\cr
}\eqno(49b)$$
Eq.~(49a) takes the form
$$S=\log Z -{\bf Tr} \tilde \lambda {\delta \log Z \over \delta \tilde \lambda}
-\hat{\tau} {\partial \log Z \over \partial \hat{\tau} }
-\tau {\partial \log Z \over \partial \tau
}~.~~~\eqno(49c)$$
Thus the entropy is a thermodynamic quantity
determined solely by the partition function.
Taking second derivatives of the partition function, we can derive
thermodynamic formulas
for the averaged mean square fluctuations of the conserved quantities
$\tilde C$, ${\bf \hat H}={\bf \hat{Tr}} H$, and ${\bf H}={\bf Tr}H$,
$$\eqalign{
\Delta_{{\bf Tr} \tilde P \tilde C}^2\equiv& \langle( {\bf Tr} \tilde P
\tilde C - \langle{\bf Tr} \tilde P \tilde C \rangle_{AV})^2 \rangle_{AV}
=\langle ({\bf Tr} \tilde P \tilde C)^2\rangle_{AV}-\langle {\bf Tr} \tilde P
\tilde C\rangle_{AV}^2 = ({\bf Tr} \tilde P {\delta \over
\delta \tilde \lambda})^2 \log Z~,~~~\cr
\Delta_{\bf \hat H}^2 \equiv&\langle ({\bf \hat H} -\langle
{\bf \hat H} \rangle_{AV})^2
\rangle_{AV}= \langle {\bf \hat H}^2 \rangle_{AV}-\langle {\bf \hat H}
\rangle_{AV}^2
={\partial^2 \log Z \over (\partial \hat{\tau})^2}~,~~~\cr
\Delta_{\bf H}^2 \equiv&\langle ({\bf H} -\langle {\bf H} \rangle_{AV})^2
\rangle_{AV}= \langle {\bf H}^2 \rangle_{AV}-\langle {\bf H} \rangle_{AV}^2
={\partial^2 \log Z \over (\partial \tau)^2}~,~~~\cr
}\eqno(49d)$$
with $\tilde P$ an arbitrary fixed anti--self--adjoint operator.
Equations (49a--d) show that the entropy, the expectations of $\tilde C$,
${\bf \hat H}$, and ${\bf H}$, and the mean square fluctuations of the latter
three
quantities,
are all extensive quantities which grow linearly with the size $N$
of the system.
This implies that the ratio of the root mean square fluctuation to the mean
for $\tilde C$, ${\bf \hat H}$, and ${\bf H}$ vanishes as $N^{-1/2}$ in the
limit
$N \to \infty$, and justifies using mean values in imposing the constraints
in Eq.~(48a).

In the Ward identity derivation of the following section, the distribution
$\rho$ enters under a phase space integral in two ways.  There are a number
of terms in which $\rho$ appears simply as a weighting factor; these terms
appear to be dominant and in them
we assume that the distribution is sharp enough so that unvaried
factors of the conserved extensive quantity
$\tilde C$ can be replaced by the ensemble average $\langle
\tilde C \rangle_{AV}$, an approximation that should
become exact in
the $N \to \infty$ limit.
In addition there is a term involving the variation $\delta \rho $ of the
equilibrium distribution, which gives corrections to the Ward identity
that we presume to come from very high energy physics.  The correction term
is evaluated using the formula
(the normalization factor Z is not varied in the following
equations, since we are interested here only in variations for which $\delta
Z=0$),
$$\eqalign{
\delta \rho =&\rho  \delta \log \rho \cr
=&\rho \left({\bf Tr} {\delta \log \rho \over \delta \tilde C} \delta \tilde C
+ {\partial \log \rho \over \partial {\bf \hat H}} \delta {\bf \hat H}
+{\partial \log \rho \over \partial {\bf H}} \delta {\bf H} \right)~~~,\cr
}\eqno(50a)$$
and from Eq.~(48c) we find for the variations of $\log \rho $,
$$\eqalign{
{\delta \log \rho \over \delta \tilde C}=&-\tilde \lambda~,~~~\cr
{\partial \log \rho \over \partial {\bf \hat H}}=&-\hat{\tau}~~~,\cr
{\partial \log \rho \over \partial {\bf H}}=&-\tau~~~.\cr
}\eqno(50b)$$
The variations $\delta_{x_s} \tilde C$,  $\delta_{x_s} {\bf \hat H}$, and
$\delta_{x_s} {\bf H}$
corresponding to the variation $\delta x_s$ are computed in
Appendix C, with the results
$$\eqalign{
\delta_{x_s} \tilde C=&\sum_r \omega_{rs} (x_r \delta x_s -\delta x_s
\epsilon_r x_r)~,~~~\cr
\delta_{x_s} {\bf \hat H} =& \delta_{x_s} {\bf \hat{Tr}} H={\bf Tr}
(-1)^F\sum_r
\dot x_r \hat \omega_{rs} \delta x_s~,~~~\cr
\delta_{x_s} {\bf H} =& \delta_{x_s} {\bf Tr} H={\bf Tr} \sum_r
\dot x_r  \omega_{rs} \delta x_s~,~~~\cr
}\eqno (50c)$$
with $\hat \omega_{rs}$ and $\alpha_{ur} \equiv
\sum_s \omega_{us} \hat \omega_{rs}$ given in
Appendix C.
In evaluating the correction term, we assume that unvaried factors of the
conserved extensive quantities ${\bf \hat H}$ and ${\bf H}$ can also be
replaced by their corresponding ensemble averages, again an approximation
that should become exact in the $N \to \infty$ limit.

As a final remark, in the derivation of the next section we shall follow
the conventional practice of introducing, for each phase space operator,
an operator source which can be varied and which is then set to zero
after all variations have been performed.  It is convenient to define the
sources $\rho_r$ so that they are all bosonic and self--adjoint.  To
couple such sources to the phase space operators, we employ the auxilliary
real or Grassmann real parameters $\sigma_r$ introduced in Sec.~2, and
couple the source term as ${\bf Tr} \rho_r \sigma_r x_r$.  When $r$ is
a bosonic index, the source $\rho_r$ takes a distinct value for each $r$,
since $\sigma_r x_r$ is already self--adjoint in this case.  When
$r$ is a fermionic index in complex quantum mechanics, the source
$\rho_r$ takes a distinct value only for each pair of fermionic
phase space operators
$q_r,~p_r$, so that each distinct $\rho_r$
multiplies the combination $\sigma_{q_r} q_r + \sigma_{p_r}
p_r$, which (remembering that $q_r^{\dagger}=-ip_r$) is self--adjoint when the
Grassmann parameters are taken to
obey $\sigma_{q_r}^{\dagger}=-i \sigma_{p_r}$.   When $r$ is a fermionic index
in real or quaternionic quantum mechanics, the source $\rho_r$ takes a
distinct value only for each quartet of fermionic
phase space operators $q_{r1},~p_{r1},
{}~q_{r2},~p_{r2}$, so that each distinct $\rho_r$ multiplies the combination
$\sigma_{q_{r1}} q_{r1} + \sigma_{p_{r1}} p_{r1} + \sigma_{q_{r2}} q_{r2}
+\sigma_{p_{r2}} p_{r2}$, which (remembering that $q_{r2}^{\dagger}=\mp
p_{r1},~p_{r2}^{\dagger}=\pm q_{r1}$) is self--adjoint when the Grassmann
parameters are taken to obey $\sigma_{q_{r2}}^{\dagger}=\pm \sigma_{p_{r1}},~
\sigma_{p_{r2}}^{\dagger}=\mp \sigma_{q_{r1}}$.  We shall follow the practice
of writing the
source term as ${\bf Tr} \sum_r \rho_r (\sigma_r x_r)$, with the parentheses
a reminder of this implicit grouping.
With the sources included, the equilibrium distribution and
partition function take the form
$$\eqalign{
\rho=&Z^{-1} \exp[-{\bf Tr} \sum_r \rho_r (\sigma_r x_r)]
\exp(-{\bf Tr} \tilde \lambda \tilde C- \hat{\tau} {\bf \hat H}-\tau
{\bf H})~~~, \cr
Z=&\int d\mu \exp[-{\bf Tr} \sum_r \rho_r (\sigma_r x_r) ]
\exp(-{\bf Tr} \tilde \lambda \tilde C-\hat{\tau} {\bf \hat H}-\tau
{\bf H})~~~. \cr
}\eqno(51a)$$
Continuing to use the expression $\langle {\cal O} \rangle_{AV}$ to denote
the average of a general operator over the equilibrium distribution of
Eq.~(51a) which includes sources, the variations of $\log Z$ with respect to
its source arguments are related to the averages of the $x_r$ by
$$\langle(\sigma_r x_r) \rangle_{AV}=-{\delta \log Z \over \delta \rho_r}
{}~.~~~\eqno(51b)$$

\bigskip
\centerline{{\bf 6.~~Ward Identities, Unitarity, and the Canonical Algebra}}

In the previous sections we have seen that in generalized quantum dynamics
there is a conserved operator $\tilde C$, given by the sum of commutators for
all of the bosonic degrees of freedom minus the corresponding sum
of fermionic anticommutators, and
that this operator plays a role in equilibrium statistical mechanics closely
analogous to that played by the summed energy of independent degrees of
freedom in
classical statistical physics.  This naturally suggests the idea
that the canonical
commutation relations of quantum mechanics may arise from a generalized
quantum dynamics analog of the classical theorem of equipartition of
energy.  To pursue this thought, let us begin by reviewing a
simple derivation [8] of the classical equipartition theorem.
Let $H(\{x_r\})$ be the classical Hamiltonian
as a function of classical phase space variables $\{x_r\}$, and let
$d\mu(\{x_r\})$ be the classical phase space integration measure.  We
consider the integral
$$\eqalign{
&\int d\mu {\partial [x_r \exp(-\beta H)] \over \partial x_s}\cr
=&\int d\mu \delta_{rs} \exp(-\beta H) \cr
-&\int d\mu x_r {\partial [\beta H] \over \partial x_s} \exp(-\beta H)~~~, \cr
}\eqno(52a)$$
the left hand side of which
is the integral of a total derivative and vanishes when
the integrand is sufficiently rapidly vanishing at infinity.  Assuming this,
we get
$$\delta_{rs}={ \int d\mu x_r \beta (\partial H/ \partial x_s)
\exp(-\beta H) \over \int d\mu \exp(-\beta H) } ~~~, \eqno(52b)$$
which is the classical theorem of equipartition of energy.  The method of
derivation is similar to that used to derive Ward identities from functional
integrals in quantum field theory (see, e.g. [9]), and the equipartition
theorem can be viewed as a Ward identity application in classical statistical
mechanics.

We proceed now to derive a Ward identity for the statistical ensemble
of generalized quantum dynamics.  The derivation is based on Eq.~(42),
which asserts
the invariance of the operator phase space measure $d \mu$ under
finite operator shifts $\delta x_r$, which can be arbitrary apart from the
obvious restriction that they must satisfy the same self--adjointness
restrictions as the corresponding operators $x_r$.  We take account of the
adjointness restrictions on the variations by using the following Lemma,
proved in Appendix D,
which insures that when we equate the variation of a total trace functional
to zero, we do not inadvertently ``deduce'' an operator relation which
arises from the variation of an anti--self--adjoint operator, which is
identically zero when acted on by {\bf Tr}.
\hfill \break
\leftline{\bf Lemma:}  Let $Y_1$ and $Y_2$ be two self--adjoint bosonic
or two anti--self--adjoint bosonic
operators constructed from the phase space variables.  Then in $0=\delta
{\bf Tr}Y_1 Y_2$, the self--adjointness restrictions on the variations can
be ignored.\hfill \break

To derive the Ward identity, we consider the expression
$$0=\int d \mu \delta_{x_s}\left[ \exp[-{\bf Tr} \tilde \lambda \tilde C
-\hat{\tau} {\bf \hat H}-\tau {\bf H} -{\bf Tr} \sum_r \rho_r (\sigma_r x_r)]
{}~~{\bf Tr} \{\tilde C, i_{eff} \} V \right]
{}~~~,\eqno(53a)$$
where the operator variation $\delta_{x_s}$ is defined to act on an arbitrary
operator $X[\{x_t\}]$ as
$$\delta_{x_s} X[\{x_t\}]= X[\{x_t,t\ne s; x_s+\delta x_s\}]
-X[\{x_t\}]~~~,\eqno(53b)$$
and where in equating the shift to zero we are using the shift invariance
of the measure and the assumption that the integrals are sufficiently
convergent that contributions from infinity can be ignored.
In Eq.~(53a), the expression $V$ denotes any self--adjoint
polynomial in the variables
$\{ (\sigma_s x_s) \}$ constructed using coefficients
which are $c$--numbers apart from a possible
dependence on the operators $(-1)^F$ and $i_{eff}$.
[Inclusion of the auxilliary factors $\sigma_s$ in the combination $(\sigma_s
x_s)$, which was defined in the discussion preceding Eq.~(51a), is
purely a matter of convenience; it permits working with bosonic quantities
throughout, and also facilitates comparison
with the form of the canonical algebra given in Eq.~(13).]
The traces in the exponent in Eq.~(53a) and the trace
involving $V$ both have the form specified by the Lemma,
so we can proceed with taking variations, giving [c.f. Eqs.~(50a, b)]
$$\eqalign{
&0=\int d \mu \exp[-{\bf Tr} \tilde \lambda \tilde C
-\hat{\tau} {\bf \hat H}-\tau {\bf H} -{\bf Tr} \sum_r \rho_r (\sigma_r x_r)]
\cr
&\times \left[ [-{\bf Tr}\tilde \lambda \delta_{x_s} \tilde C
- \hat{\tau} \delta_{x_s} {\bf \hat H} -\tau \delta_{x_s} {\bf H}
-{\bf Tr}  \rho_s \sigma_s \delta x_s ]
{\bf Tr}\{\tilde C, i_{eff} \} V  \right. \cr
&+ \left. {\bf Tr}( \{\delta_{x_s} \tilde C, i_{eff} \} V
+\{ \tilde C,i_{eff}\} \delta_{x_s} V)\right]
{}~~.\cr
}\eqno(53c)$$

We now make two assumptions: First, when the extensive
quantity $\tilde C$ appears in {\it unvaried} form as a factor in
an ensemble average over the equilibrium distribution, we assume
that it
can be replaced by its ensemble average  $\langle \tilde C \rangle_{AV}$
(but this is
{\it not}, of course, applied to the $\tilde C$ in the exponent of the
equilibrium distribution).  This assumption amounts to neglecting the
fluctuations of $\tilde C$ when it appears as a factor
in the integrand, and as we argued in Sec.~5, should be justified
in the limit $N \to \infty$. Second, we assume the
form of Eq.~(46b) for $\langle \tilde C \rangle_{AV}$.  Replacing unvaried
factors of $\tilde C$ by their ensemble averages, substituting
Eqs.~(46b) and (50c), and making some cyclic permutations under the graded
trace, Eq.~(53c) becomes
$$\eqalign{
0=&\int d \mu \exp[-{\bf Tr} \tilde \lambda \tilde C
-\hat{\tau} {\bf \hat H} -\tau {\bf H} -{\bf Tr}
\sum_r \rho_r (\sigma_r x_r)] \cr
\times &~~\left[ \left(-{\bf Tr} [\tilde \lambda,\sum_r \omega_{rs} x_r]
\delta x_s- \hat{\tau} {\bf Tr}(-1)^F\sum_r \dot x_r \hat \omega_{rs} \delta
x_s
\right.  \right. \cr
-&\left. \tau {\bf Tr} \sum_r \dot x_r \omega_{rs} \delta x_s
-{\bf Tr}  \rho_s \sigma_s \delta x_s  \right)
(-2 \hbar){\bf V}   \cr
+& \left.{\bf Tr} [ \{i_{eff}, V \}, \sum_r \omega_{rs} x_r] \delta x_s
-2 \hbar {\bf Tr} {\delta {\bf V} \over \delta x_s} \delta x_s \right]
{}~~~,\cr
}\eqno(54a)$$
where in the final line we have used the definition of the operator
derivative of the total trace functional ${\bf V}={\bf Tr}V$,
$${\bf Tr} \delta_{x_s} V=
{\bf Tr} {\delta {\bf V} \over \delta x_s} \delta x_s~~~.\eqno(54b)$$

Before proceeding further, let us examine the structure of the first term on
the right hand side of Eq.~(54a), which contains the factor
$[\tilde \lambda,\sum_r \omega_{rs} x_r]$.  After varying with respect to
the sources and setting the sources to zero, this term is proportional to
$${\bf Tr} [\tilde \lambda, \int d\mu \rho P(\{ x_r \}) {\bf V} ]
\delta x_s~,~~~
\eqno(54c)$$
with $\rho$ the zero source equilibrium distribution of Eq.~(48c) and with
$P$ the polynomial which results after variation with respect to the sources.
Since $V$, by assumption, involves no operator coefficients other
than $(-1)^F$ and
$i_{eff}$, the reasoning of Eqs.~(45a--c) implies that the ensemble
average in Eq.~(54c) is a function $G(\tilde \lambda,(-1)^F)$, and so commutes
with $\tilde \lambda$.  Hence the first term on the right hand side of
Eq.~(54a)
vanishes.  Since each remaining term in Eq.~(54a) is the graded
trace of an operator times the
variation  $\delta x_s$, we can equate the total operator coefficient
of this variation to zero.  After multiplying through by ${1 \over 2} \sum_s
\omega_{us}$, and using Eq.~(10c) and the definition of $\alpha_{ur}$
following Eq.~(50c),
this gives the operator Ward (or equipartition) identity
$$\eqalign{
0=&\int d \mu \exp[-{\bf Tr} \tilde \lambda \tilde C
-\hat{\tau} {\bf \hat H}-\tau {\bf H}-{\bf Tr} \sum_r \rho_r (\sigma_r x_r)]
\cr
\times& ~~\left[ [\hat{\tau}(-1)^F\sum_{r} \alpha_{ur} \dot x_r
+\tau\dot x_u+\sum_s \omega_{us} \sigma_s \rho_s ] \hbar{\bf V}
\right. \cr
+& \left. [{1 \over 2} \{i_{eff}, V \}, x_u]
- \hbar \sum_s \omega_{us} {\delta {\bf V} \over \delta x_s}  \right]
{}~~~.\cr
}\eqno(55a)$$
Dividing by the partition function $Z$, Eq.~(55a) can be rewritten in the
compact form
$$0=\langle
[\hat{\tau}(-1)^F\sum_{r} \alpha_{ur} \dot x_r
+\tau \dot x_u +\sum_s \omega_{us} \sigma_s \rho_s ] \hbar {\bf V}
+ [{1 \over 2} \{i_{eff}, V \}, x_u]
- \hbar \sum_s \omega_{us} {\delta {\bf V} \over \delta x_s}
\rangle_{AV}~,\hfill \break~~~\eqno(55b)$$
with $\langle ~~~\rangle_{AV}$ denoting the ensemble
average with sources present, and
with the understanding that after variation with respect to the sources, the
sources
are to be set equal to zero.  A detailed discussion of how symmetrized
polynomials in the phase space operators may be built up through source
variation is given in Appendix E, and a discussion of a second Ward identity
connected to the conserved operator $\tilde F$ is given in Appendix G.
Although the explicit
source term in Eq.~(55b) proportional to $\sum_s \omega_{us} \sigma_s \rho_s$
does contribute when varied with respect to a source
$\rho_{v}$ for which $\omega_{uv}$ is nonzero, the fact that
$\omega_{uv}$ is antisymmetric in bosonic indices, and symmetric in fermionic
indices, implies that this term drops out of Weyl ordered expressions in
$x_u$ and the additional factors of $x_v,...$ brought down by
source variation, as is discussed in more detail in Appendix E.
Hence we shall drop this term in the Ward identity
applications that follow.  A closely related remark is that
although Eq.~(55b)
makes a statement of effective equality (when the $\hat{\tau}$ and $\tau$
terms are dropped) between the commutator expression in the next to last
term and the total trace derivative in the final term, this does not contradict
the counterexample of Eqs.~(17a, b) above, since the Ward identity does {\it
not}
imply that the commutator can be evaluated in terms of the canonical algebra
by using the Leibnitz product rule.  Use of the Leibnitz product rule
for $V$ is justified in Appendix E only when $V$ can be constructed as a Weyl
ordered polynomial in the operator phase space variables $\{ x_r \}$.

We proceed now to give three applications of Eq.~(55b).
As our first application
we choose $V$ to be the operator Hamiltonian $H$, so that ${\bf V}$ becomes
the conserved quantity ${\bf H}$.  Substituting the generalized quantum
dynamics equation of motion of Eq.~(9b), the Ward identity becomes
$$0=\langle
[\hat{\tau} (-1)^F\sum_{r} \alpha_{ur} \dot x_r +\tau
\dot x_u] ~\hbar {\bf H}
+ [{1 \over 2} \{i_{eff}, H \}, x_u]
- \hbar \dot x_u  \rangle_{AV}~.~~~\eqno(56a)$$
We can simplify Eq.~(56a) considerably by noting that
since ${\bf H}$ is a conserved extensive quantity, in the large $N$ limit we
can approximate it by
its ensemble average $\langle {\bf H} \rangle_{AV}$;  its coefficient in
Eq.~(56a) is then proportional to the partition function variation
$\sum_s \omega_{us} \delta Z / \delta x_s =0.$
So dropping the $\hat{\tau} $ and
$\tau$ terms, and
multiplying Eq.~(56a) by $-1$ we are then left with
$$0=\langle \hbar \dot x_u - [{1 \over 2}  \{i_{eff} , H \} , x_u]
\rangle_{AV}~~~,\eqno(56b)$$
where the ensemble used to form the average is understood to still contain
nonzero sources.

It is convenient at this point to recall the properties of $i_{eff}$
given in Eq.~(46a), and to make the definition
$$\tilde H_{eff} \equiv {1 \over 2}  \{ i_{eff}, H \}~,~~
\tilde H_{eff}^{\dagger}=-\tilde H_{eff} ~~~, \eqno(57a)$$
with the tilde indicating that $\tilde H_{eff}$
is an anti--self--adjoint operator.  We find from Eq.~(57a) that
$$i_{eff} \tilde H_{eff} = {1 \over 2} (-H + i_{eff} H i_{eff} )
=\tilde H_{eff} i_{eff}~~~,\eqno(57b)$$
in other words, $i_{eff}$ and $\tilde H_{eff}$ commute.
We can now write Eq.~(56b) as
$$0=\langle \hbar \dot x_u -   [\tilde H_{eff} , x_u]
\rangle_{AV}~~~,\eqno(57c)$$
which is an effective Heisenberg picture equation of motion for $x_u$ in
anti--self--adjoint generator form.
Let us now vary with respect to the sources, leading (as described in
Appendix E) to the replacement of Eq.~(57c) by the expression
$$0=\langle \hbar \dot  P(\{ x_r \})
-  [\tilde H_{eff} , P(\{ x_r \})]
\rangle_{AV}~~~,\eqno(57d)$$
with $P(\{ x_r \})$ a Weyl ordered polynomial
formed with coefficients which are $c$--numbers (apart from a possible
dependence on $(-1)^F$ and
$i_{eff}$).  In particular, letting $P$ be the effective
Hamiltonian $\tilde H_{eff}$, we find that
$${d \over dt} \langle \tilde H_{eff}\rangle_{AV}=0~~~,\eqno(57e)$$
and so $\langle \tilde H_{eff} \rangle_{AV}$, still in the presence of
sources, is a constant of the motion.  Thus the ensemble averages of products
of the coordinates have an effective unitary time development, involving
an anti--self--adjoint effective time independent Hamiltonian.

This time
development, however, cannot immediately be put into the standard Heisenberg
picture form of Eq.~(16b), which involves a self--adjoint Hamiltonian.
We now show that we can extract from the $\{ x_r \}$ a set of new operators
$\{ x_{r~eff} \}$, which do obey an effective dynamics of the
standard Heisenberg form.   We begin by introducing the
self--adjoint effective Hamiltonian $H_{eff}$ defined by
$$H_{eff} = -i_{eff} \tilde H_{eff} = {1 \over 2} (H - i_{eff} H i_{eff})
{}~~~,\eqno(58a)$$
which evidently also commutes with $i_{eff}$.
In analogy with Eq.~(58a), we further define
$$x_{r~eff}={1\over 2} (x_r - i_{eff} x_r i_{eff})~~~,\eqno(58b)$$
which obeys
$$i_{eff} x_{r~eff} = {1  \over 2} (i_{eff} x_r +x_r i_{eff})=x_{r~eff} i_{eff}
{}~~~,\eqno(58c)$$
and thus also commutes with $i_{eff}$.  For any operators $x_1,x_2$  this
definition evidently obeys
$$(x_1 x_{2eff})_{eff}=(x_{1eff}x_2)_{eff}=x_{1eff}x_{2eff}~~~,\eqno(58d)$$
and so taking the effective projection of the equation of motion of Eq.~(57c)
gives
$$0=\langle \hbar \dot x_{u~eff} -   [\tilde H_{eff} , x_{u~eff}]
\rangle_{AV}~~~,\eqno(59a)$$
which by Eq.~(58c) can now be written directly in terms of the
self--adjoint effective Hamiltonian defined in Eq.~(58a),
$$0=\langle \hbar \dot x_{u~eff} -i_{eff}   [H_{eff} , x_{u~eff}]
\rangle_{AV}~~~,\eqno(59b)$$
and so has the standard Heisenberg picture form of complex quantum
mechanics.  In a similar fashion, by repeated applications of Eq.~(58c) to
polynomials of successively one higher degree, we can derive a
self--adjoint analog of Eq.~(57d),
$$0=\langle \hbar \dot  P(\{ x_{r~eff} \}) -i_{eff}
[ H_{eff} , P(\{ x_{r~eff} \})]
\rangle_{AV}~~~.\eqno(59c)$$

Before proceeding to further Ward identity applications, we make two remarks.
The first is that the method of projecting out effective operators which
commute with $i_{eff}$ is simply a complex analog of the method
of extracting ``formally real''components of operators in quaternionic
quantum mechanics [2].  The second is that the
operators $x_r$ and $x_{r~eff}$ always differ in the cases of real and
quaternionic Hilbert spaces. Even in the case of a complex Hilbert space
they differ when $i_{eff} \not= i$, as shown in Appendix B, while in Appendix
G we show that the complex case with $i_{eff}=i$ is excluded.
When $x_r$ and $x_{r~eff}$ differ, there is a ``hidden sector''
which cannot be attained by acting with arbitrary polynomials formed from
the effective operators alone.

We turn now to the second Ward identity application, which we derive by
taking $V$ in Eq.~(55b) to be $V=(\sigma_t x_t)$. The parentheses here
indicate summation over the one, two, or four $t$ values corresponding to a
distinct source term $\rho_t$ in the equilibrium distribution, which according
to the discussion preceding Eq.~(51a) makes $V$ self--adjoint.  Since the
$\sigma$ parameters which appear in this sum are linearly independent, we
can ignore the parentheses and simply substitute the single term
$V=\sigma_t x_t$ into Eq.~(55b), corresponding to
$ \delta {\bf V}/\delta x_s= \sigma_t\delta_{ts}$, giving
$$0=\langle
[\hat{\tau}(-1)^F\sum_{r} \alpha_{ur} \dot x_r +\tau
\dot x_u] ~\hbar{\bf Tr}
\sigma_t x_t  +[{1 \over 2} \{ i_{eff}, \sigma_t x_t \}, x_u]
-\hbar \omega_{ut} \sigma_t ~\rangle_{AV}~~~.\eqno(60a)$$
Referring back to Eqs.~(58a--d), we see that Eq.~(60a) can be rewritten in
terms of $x_{t~eff}$ as
$$0=\langle
[\hat{\tau}(-1)^F\sum_{r} \alpha_{ur} \dot x_r +\tau
\dot x_u]
\hbar{\bf Tr}
\sigma_t x_{t~eff}  +[i_{eff} \sigma_t x_{t~eff}, x_u]
-\hbar \omega_{ut} \sigma_t ~\rangle_{AV}~~~.\eqno(60b)$$
Taking the effective projection of Eq.~(60b) by using Eq.~(58d), and then
multiplying through by $i_{eff}$, we get finally
$$0=\langle
i_{eff}[\hat{\tau}(-1)^F\sum_{r} \alpha_{ur} \dot x_{r~eff}
+\tau \dot x_{u~eff}]
\hbar{\bf Tr} \sigma_t x_{t~eff}
+[x_{u~eff}, \sigma_t x_{t~eff}]
-i_{eff} \hbar \omega_{ut} \sigma_t ~\rangle_{AV}~~~,\eqno(60c)$$
which coincides in form with the canonical algebra of Eq.~(13) (where
we had set $\hbar=1$) when the dynamics dependent terms
proportional to $\hat{\tau}$  and $\tau$
are dropped.  Since these two terms
are proportional to time derivatives $\dot x_{eff}$, their neglect should be
justified
in a regime characterized by energies much lower than the energy scale of the
underlying dynamics.  (Also, since these two terms are proportional to the
graded trace ${\bf Tr} \sigma_t x_{t~eff}$, they may be further suppressed
by boson fermion cancellation.)
By repeated applications of Eq.~(58d) to the cases in which $V$ is a
successively one degree higher self--adjoint
polynomial, one proves similarly from Eq.~(55b) that
after the $\hat{\tau}$ and $\tau$ terms are dropped,
$$0=\langle~[x_{u~eff},  V(\{ x_{r~eff} \})~]
- i_{eff}\hbar \sum_s\omega_{us} {\delta {\bf V}(\{ x_{r~eff} \})
\over \delta x_{s~eff}} ~\rangle_{AV}~~~,\eqno(60d)$$
corresponding in form (when $\hbar$=1)
to the canonical algebra expression comprising the
second and third terms of Eq.~(29).  Similarly, as discussed in Appendix E,
by varying the sources in Eq.~(60b) one can justify (when V is Weyl ordered)
the evaluation of the commutator in Eq.~(60d) in terms of the canonical
algebra of Eq.~(13) using the Leibnitz product rule.

As a final application of Eq.~(55b), we examine its implications for
canonical transformations.   Replacing $V$ now by $G$, with {\bf G} the
generator of a Weyl ordered intrinsic canonical transformation, comparing
with Eq.~(27) and dividing by $\hbar$, and dropping the $\hat{\tau}$ and
$\tau$
terms,
we see that Eq.~(55b) takes the form
$$0=\langle~
  [ {1\over 2}\hbar^{-1} \{i_{eff},G\}, x_u]-\delta x_u ~\rangle_{AV}~~~.
\eqno(61a)$$
Defining the anti--self--adjoint generator $\tilde G_{eff}$ by
$$\tilde G_{eff} = {1 \over 2} \{i_{eff},G\}~~~, \eqno(61b)$$
Eq.~(61a) can be rewritten in the form
$$0=\langle~  \delta x_u-\hbar^{-1} [\tilde G_{eff}, x_u]
\rangle_{AV}~~~,
\eqno(61c)$$
which after taking the effective projection gives
$$0=\langle~  \delta x_{u~eff}-\hbar^{-1} [\tilde G_{eff}, x_{u~eff} ]
\rangle_{AV}~~~,
\eqno(61d)$$
and corresponds in form (when $\hbar=1$) to the first
and third terms of Eq.~(29).
Equations (61c, d), and their Weyl ordered polynomial generalizations
$$\eqalign{
0=&\langle~  \delta P(\{ x_r \})-\hbar^{-1} [\tilde G_{eff}, P(\{ x_r \}) ]
\rangle_{AV}~~~ ,\cr
0=&\langle~  \delta P(\{ x_{r~eff} \})-\hbar^{-1}
[\tilde G_{eff}, P(\{ x_{r~eff} \})]
\rangle_{AV}~~~ ,\cr
}\eqno(61e)$$
which are analogous to Eqs.~(57d) and (59c),
indicate that there is a correspondence
between Weyl ordered intrinsic
canonical transformations in the underlying generalized
quantum dynamics, and unitary transformations $U_{eff}$ of the form
$$U_{eff}=\exp(\hbar^{-1} \tilde G_{eff}) \eqno(62)$$
acting on the variables $\{ x_{r~eff} \}$.  The invariance of $i_{eff}$
under unitary transformation by $U_{eff}$ is an image, in the effective
theory, of the invariance of $\tilde C$ under intrinsic
canonical transformations in
the underlying generalized quantum dynamics.

To summarize, we have shown that there is a striking correspondence between
the structure of the set of ensemble averages calculated in generalized
quantum dynamics and the structure of canonical quantum field theory. To
what specific field theoretic structures do these averages correspond?  To
answer this question we note that in the absence of sources,
the averages of monomials constructed from the phase space operators all are
functions, constructed with real number coefficients,
solely of the operator $i_{eff}$ and of the grading operator $(-1)^F$. (This
statement follows from an argument given at the end of Appendix B.)
Since the Hamiltonian $H$
is necessarily bosonic, physical processes cannot
change the value of $(-1)^F$, and so
the $(-1)^F=1$ and $(-1)^F=-1$ sectors of Hilbert space
are superselection sectors, the states of which do not superimpose.
Within each superselection sector, the ensemble averages of monomials
are simply linear combinations with real coefficients
of the two operators 1 and $i_{eff}$; this implies
that they are to be indentified with matrix
elements, rather than operators, in an effective complex field theory
in which $i_{eff}$ plays the role of the imaginary unit.
Since for any self--adjoint operator $A$ formed from the phase space operators
$\{ x_r \}$, reality and positivity of the equilibrium phase space density
$\rho$ imply that $\langle A \rangle_{AV}$ is real and $\langle A^{\dagger}
A \rangle_{AV}$ is nonnegative, the ensemble averages must correspond to
expectation values in some pure or mixed state in the effective field theory.
But since the ensemble in which the averages are formed is Lorentz invariant
(recall that in a general Lorentz frame ${\bf \hat H}$ gets replaced by the
invariant $[-({\bf  \hat{Tr}}\,p_{\mu})^2]^{1\over2}$, and similarly for ${\bf
H}$),
and since the vacuum is the only
Lorentz invariant state in quantum field theory,
it is then natural to identify the ensemble averages with vacuum
expectation values in an effective quantum field theory.
This leads us to conjecture that in the limit of infinitely many degrees of
freedom and an infinite dimensional underlying Hilbert space,
{\it the ensemble averages of Weyl ordered
polynomials formed from the
canonical phase space operators in generalized quantum dynamics are
isomorphic in structure to the vacuum expectation values of the corresponding
Weyl ordered polynomials formed from the canonical operators in complex
quantum field theory}.

An interesting feature that has emerged from our calculations
is that an effective complex structure results irrespective of whether one
starts from an underlying real, complex, or quaternionic Hilbert space.
In particular, there now seems to
be no reason to exclude the aesthetically appealing case of real Hilbert
space, since we have given a natural and automatic mechanism for the
complexification which is needed to describe the observed physical world.
We also note that there is a natural connection between our calculations
and matrix models containing $2N$ matrices, which  correspond to our $2N$
phase space operators, acting on an $M$ dimensional space of states in the
large $M$  limit (conventionally called the ``large $N$'' limit in the
matrix model literature).  When the term in the equilibrium ensemble
containing $\tilde \lambda$  vanishes, the ensemble is
invariant under unitary transformations,
in which case the large $M$ limit is known to be described by the
{\it classical} field theory [10, 11, 12] of a so--called ``master field''.
When the $\tilde \lambda$ term is  nontrivial
it breaks the unitary invariance, raising the interesting possibility that
{\it the master field in this case is a complex quantum field}.

We make next some observations concerning the use of Weyl
ordering in our analysis.  It is clear from the discussion
of Secs.~2--4 and Appendix E that the proposed isomorphism must
fail for certain types of polynomials
which are not Weyl ordered.  There is evidently a subtlety
in the non--Weyl ordered case, an understanding of which will require a
more detailed investigation. Since the operator gauge invariant models
discussed in Refs. [1, 2] do not have Weyl ordered Hamiltonians, it is
important to extend our derivations so that this case (which involves
fourth and lower degree polynomials in the phase space operators )
is explicitly included; this issue will be addressed in a
separate publication.

Finally, we remark that in generalized quantum dynamics the concept of
``temperature'' is not defined, because the operator Hamiltonian $H$ is
not a constant of the motion: only its graded and
(in the class of models studied here)
ungraded traces
${\bf Tr} H$ and ${\bf \hat{Tr}} H$ are
conserved.  However, in the effective complex theory the effective
Hamiltonian is conserved, making possible further fine grained
equilibria governed by $H_{eff}$, and thus leading
to the emergence of the standard thermal ensemble $\exp(-\beta H_{eff})$
parameterized by the temperature $\beta^{-1}$.

\bigskip
\centerline{{\bf 7.~~Conditions for Convergence}}

The Ward identities derived in the preceding section are meaningful only when
the phase space integrals which appear in them are convergent.  As we have
already noted, the generic conserved quantities $\tilde C$ and ${\bf H}$ are
both indefinite in sign, and so when standard statistical mechanical methods
are applied to them one finds a partition function which is divergent.  A
necessary condition for convergence is that there be at least one constant
of motion which is bounded below, and an obvious candidate for this is the
ungraded trace ${\bf \hat{Tr}}H$.  A simple criterion can be given for
constructing
Lagrangians, in which the total trace equations of motion obtained using the
graded trace Lagrangian ${\bf Tr}L$ imply conservation of
${\bf \hat{Tr}}H$ as well as conservation of
${\bf Tr}H$.  Let us consider Lagrangians in which the
fermion fields appear only through one of the standard bilinears of the
form $\psi_l^{\dagger}...\psi_r$; then when bosonic variables to which the
fermions couple are ordered to the {\it outside} of the fermion factors, as
in the vector coupling $\psi_l^{\dagger} \gamma^0 \gamma^{\mu} ...\psi_r
B_{\mu}$ or the
scalar coupling $\psi_l^{\dagger} \gamma^0 ...\psi_r B$, the ungraded
trace and graded
trace equations of motion are the same and imply conservation of
${\bf \hat{Tr}}H$.
The reason is that when the bosons are ordered to the outside of the fermions,
only cyclic permutations of bosonic factors are required to construct the
equations of motion for the bosons and for $\psi_r$, and so these are the
same irrespective of whether or not a grading factor is included inside the
trace.  The equation of motion for $\psi_l^{\dagger}$ can
then be obtained as the
adjoint of one of the $\psi_r$ equations of motion, or if calculated directly
by permuting the factor $\delta \psi_l^{\dagger}$ to
the right, the same grading factor
appears in all terms and so drops out of the $\psi_l$ equation of motion.
Closer examination shows that in the models with a conserved ${\bf \hat{Tr}}H$,
the
equations of motion also imply conservation of an ungauged fermion number
current $\tilde F^{\mu}$, with $\int d^3x \tilde F^0$ yielding
the conserved charge $\tilde F$ discussed
briefly in Sec. 5 and in Appendices C and G.

On the other hand, if the bosonic variables to which the fermions couple are
ordered to the {\it inside} of the fermion factors, as in the vector coupling
$\psi_l^{\dagger} \gamma^0 \gamma^{\mu} B_{\mu} \psi_r$, then $\psi_r$
must be cyclically
permuted to the left in order to obtain the $B_{\mu}$ equation of motion,
resulting in differing signs for the $B_{\mu}$ source
terms in the graded and ungraded
cases.  These features of the generalized quantum dynamics equations of motion
are readily verified by examination of the catalog of models given in
Secs.~13.6--13.7 of [2].  In generalizing the Lagrangians for the various
gauge theory and Higgs components of the standard model, and its grand unified
extensions, so as to give generalized quantum dynamics Lagrangians,
one is always free to adopt the ordering convention which leads to
conservation of ${\bf \hat{Tr}}H$.   However, one can readily construct
generalized
quantum dynamics models in which ${\bf \hat{Tr}}H$ is not conserved; an example
is
the maximally operator gauge invariant model constructed in [1, 2] using
two fermion
fields, in which the gauge potential $B_{\mu}^{\prime}$ is ordered to the
outside of the fermion fields and a second gauge potential $B_{\mu}$ is
ordered to the inside, the difference in ordering being precisely what
distinguishes the action of the two gaugings.  In this case, one readily
verifies that the source term for $B_{\mu}$ changes sign as one goes from the
graded to the ungraded equations of motion, while that for $B_{\mu}^{\prime}$
has the same sign in both cases.

In addition to the requirement that ${\bf \hat{Tr}}H$ should be conserved,
it is also necessary that $\hat{\tau} {\bf \hat{Tr}}H$ should be bounded below
and
should dominate over the indefinite
terms ${\bf Tr} \tilde \lambda \tilde C$ and $\tau {\bf H}$,
in order for the partition function
to converge.  In general, ${\bf \hat{Tr}}H$ contains three types of terms:
bosonic
kinetic energy terms, fermionic kinetic energy terms, and potential terms.
The bosonic kinetic energy terms in standard field theory models are always
positive, and we believe that in many models one will be able
to establish that the potential is bounded below.
However, because of negative energy states the fermionic kinetic
energy terms
are problematic.  For a single Dirac fermion, before any reorderings of
field operators are done, the kinetic energy (including the mass term)
has the form
$$H_{\rm kin}=\sum_{\vec p, s}(\vec p^{~2} +m^2)^{1\over 2}
(b^{\dagger}_{\vec p,s} b_{\vec p,s}
-d_{\vec p,s} d^{\dagger}_{\vec p,s})~,~~~\eqno(62a)$$
with  $m$ the fermion mass.  Since Eq.~(62a) is the difference of two
positive semidefinite terms, ${\bf \hat{Tr}}H_{\rm kin}$ is unbounded below
when the operators $b_{\vec p,s}$ and
$d_{\vec p,s}$ are independent.  However, for a self--conjugate Majorana
fermion field one has $d_{\vec p,s}=b^T_{\vec p,s}$, with $T$ the operator
transpose, and so in this case the trace of the second term in Eq.~(62a) is
$${\rm Tr}d_{\vec p,s} d^{\dagger}_{\vec p,s}=
{\rm Tr}b^T_{\vec p,s} b^{\dagger ~T}_{\vec p,s}=
{\rm Tr}b^{\dagger}_{\vec p,s} b_{\vec p,s}~,~~~\eqno(62b)$$
and cancels the trace of the first term; thus the trace of the kinetic
energy term
vanishes and the negative energy catastrophe is avoided.  Hence in order for
a field theoretic model to be extendable to a generalized quantum dynamics
model with a convergent partition function, one must be able to rewrite
its fermion kinetic energy terms entirely in terms of Majorana fermions,
which requires that every chiral fermion be accompanied by an opposite
chirality partner.   To complete the discussion of the fermion kinetic
energy terms, let us examine the behavior of the graded trace of the
fermion kinetic energy.  For any two operators ${\cal O}_1$ and ${\cal O}_2$,
irrespective of their grade, we have
$$\eqalign{
{\bf Tr} {\cal O}^T_1 {\cal O}^T_2=& {\rm Re}\sum_{m,n} \epsilon_m
{\cal O}^T_{1mn} {\cal O}^T_{2nm}={\rm Re}\sum_{m,n} \epsilon_m
{\cal O}_{1nm} {\cal O}_{2mn} \cr
=&{\rm Re}\sum_{m,n} \epsilon_m {\cal O}_{2mn}
{\cal O}_{1nm}={\bf Tr} {\cal O}_2 {\cal O}_1~~~;\cr
}\eqno(62c)$$
hence the rearrangement of Eq.~(62b) is also valid inside ${\bf Tr}$, and
so for Majorana fermions the graded trace of the kinetic energy term also
vanishes.  [The difference between Eq.~(62c) and the cyclic identity
of Eq.~(2) is
that to interchange the untransposed operators one must replace $\epsilon_m$
by $\epsilon_n$, and the product $\epsilon_m \epsilon_n$ is just the grade
of the operators ${\cal O}_{1,2}$, leading to the $\pm$ sign in Eq.~(2).]

Although a discussion of sufficient conditions for convergence of the
partition function is beyond the scope of this paper, let us give one
example which suggests that convergence can be attained.  Let us work in
complex Hilbert space, and consider special simple choices for the
ensemble parameter $\tilde \lambda$.  The simplest possibility,
$\tilde \lambda = i R$, with $R$ a positive real number, is
inadmissible because $i$ is
a $c$--number and so ${\bf Tr} \tilde \lambda \tilde C$ reduces to a
multiple of ${\bf Tr} \tilde C$, which is identically zero,
$${\bf Tr}\tilde C={\bf Tr}\sum_{r, s}x_r \omega_{rs} x_s=
{\bf Tr} \sum_{r, s}x_s \epsilon_s \omega_{rs} x_r
=-{\bf Tr}\sum_{s, r} x_s \omega_{sr} x_r =0~~~.\eqno(63a)$$
So let us consider the next simplest case, which is $\tilde \lambda=
(-1)^FiR$, for which
$$\eqalign{
{\bf Tr}\tilde \lambda C=&{\rm Re}~iR {\rm Tr}\tilde C \cr
=&{\rm Re}~iR {\rm Tr}(\sum_{r,B}[q_r,p_r]-\sum_{r,F}\{q_r,p_r\}) \cr
=&-{\rm Re}~iR{\rm Tr}\sum_{r,F}\{q_r,p_r\}=R{\rm
Tr}\sum_{r,F}\{q_r,q_r^{\dagger}\}
{}~,~~~\cr
}\eqno(63b)$$
where in the final line we have assumed that all fermions have been constructed
with the standard adjointness assignment $p_r=iq_r^{\dagger}$.
Since both terms in the anticommutator on the right hand side
of Eq.~(63b) are positive semidefinite, the term ${\bf Tr} \tilde
\lambda \tilde C$ is bounded below in this case, and does not have to be
dominated by the energy term $\hat{\tau} {\bf \hat{Tr}} H$.  So a model in
which
the ungraded trace energy is conserved, bounded below, and becomes positive
infinite on all paths approaching infinity in operator phase space (which our
preliminary investigations suggest may be possible for gauge theories)
then gives a convergent partition function.  Although in this special case the
$\tilde \lambda $  factor in the exponent has
decoupled from the bosons, as long as the bosons and
fermions interact the breaking of unitary transformation invariance implied
by the presence of $\tilde \lambda$ is still
felt by the bosons, and the model is satisfactory.  For a noninteracting
theory, of course, the fermionic and bosonic contributions to $\tilde C$
are separately conserved; both must then be included in the statistical
mechanical equilibrium distribution, and the choice $\tilde \lambda=(-1)^FiR$
would
then lead to classical mechanics, rather than quantum mechanics,
for expectations defined within the bosonic ensemble.
\bigskip
\centerline{\bf 8.~~Discussion}

In the foregoing, we have given strong evidence
that the canonical quantization rules,
which are the basis of conventional quantum mechanics and quantum field
theory, can arise as an emergent property in a generalized quantum operator
dynamics when statistical mechanical methods are applied to the ensemble
of operator initial values.  In models where the index $r$ is a composite
index composed of a spatial coordinate $\vec x$ as well as a discrete field
index, the emergent canonical algebra implies locality, even though the
underlying generalized quantum dynamics is highly nonlocal.
To our knowledge, this is the first time that
an embedding of quantum mechanics in a larger structure has been achieved
that applies to local relativistic quantum field theories.

An interesting feature of the Ward identity applications of Sec.~6 is that
the derivation of a unitary effective dynamics requires a less stringent
approximation (the replacement of ${\bf H}$ by its ensemble average)
than does the derivation of the canonical algebra (where a ``low''
frequency approximation is needed).  Thus, there may be a high energy
domain where the
statistical mechanical analysis is still valid and takes the form of a
nonlocal but unitary complex effective field dynamics, such as a string--type
theory.  This effective theory would still be only a statistical approximation
to the fully nonlocal underlying generalized quantum dynamics, and its
structure may well prove more intractable than that of the underlying theory.

An exciting
aspect of our construction is that the principal features of quantum
mechanics basically become statements about the geometry of matrices.  Thus,
if in labeling the rows and columns of matrix operators one orders all the
bosonic states before all the fermionic states, then the distinction between
bosonic and fermionic operators is simply that between matrices that are
block diagonal, and ones that are block skew diagonal, respectively.
Furthermore, as we have seen
in Sec. 6, the complex structure, canonical algebra, and unitary dynamics of
quantum mechanics are all reflections of the cyclic invariance of
the trace, which is both the origin of the conserved operator $\tilde C$
and the basis for erecting a generalized dynamics on noncommutative phase
space.  These observations suggest that the distinction between matter
degrees of freedom on the one hand, and gravitational or metric degrees of
freedom on the other, may be similarly rooted in some simple geometric
property of generalized quantum dynamics.

If our conjectured isomorphism can be proved, and if a generalized quantum
dynamics underlies the observed universe, there will be profound implications
for some of the vexing issues in conventional quantum mechanics.  One of these
issues is the quantum measurement problem.  In the
underlying generalized dynamics there are no ``dice'': the underlying dynamics
is a generalization of classical mechanics to noncommuting phase space
operators and is deterministic, although not in general unitary.
However, the ability to follow this
deterministic evolution in detail is lost at
the level of the statistical ensemble average, where a unitary
conventional quantum mechanics emerges.  In this picture, a calculation
of corrections to
the ensemble average approximation should permit the resolution of
the troubling ``paradoxes'' of quantum measurement theory.  A second issue
where our picture has important ramifications is the problem of infinities in
quantum field theory.  These divergences arise, fundamentally, because of the
singularity of the local canonical commutator/anticommutator structure of field
theory, which we have argued is an emergent property
of ensemble averages in generalized quantum dynamics.  The underlying
dynamics is nonlocal and nonsingular, and should give finite
answers to physical calculations.

There are clearly a number of important questions that must be addressed
in future work.  One of them is to give a detailed justification of the
statistical mechanical aspects of our calculation, including a
complete classification of Hamiltonians for which one can prove convergence
of the partition function, a proof that a thermodynamic limit exists,
and a justification of the assumption that
each unvaried factor of $\tilde C$ (and of other conserved
extensive quantities) in the Ward identity can be replaced
by the corresponding
ensemble average.  The emergence of quantum mechanics from the Ward identity
required the neglect
of the $\hat{\tau}$ and $\tau$ terms,
which may be reasonable only in models that have
a very large ratio of the high mass
scale characterizing pre--quantum mechanical physics to the low mass scale
characterizing quantum physics, and possibly also
a high degree of boson--fermion symmetry as well.
Thus, finding a model in which
neglect of the dynamics dependent
terms can be justified may be tantamount to finding a model that solves the
``hierarchy problem'' of explaining
the extraordinarily large ratio between the Planck mass scale
and the standard model mass scale in particle physics.
Since the entire generalized quantum dynamics formalism
seems to naturally invite
the incorporation of boson--fermion symmetries, such as
generalized forms of supersymmetry, it will be important to analyze such
symmetries.  A possibly related question is to determine
what happens when the numbers $N_B,~ N_F$
of bosonic and fermionic degrees of freedom are equal.  This issue arises
because if it were evaluated by restriction to the canonical algebra,
the conserved operator $\tilde C$ would be
equal (with $\hbar$=1) to $i_{eff}(N_B-N_F)$, which vanishes when $N_B=N_F$.
This suggests that the conditions for validity of our analysis
may be particularly delicate -- and interesting -- in models with a high
degree of boson--fermion symmetry.
Yet another question is what happens in constrained
theories, such as the operator gauge invariant theories
formulated in Refs. [1, 2].  In canonical gauges, where the constraints
can be explicitly eliminated, our
analysis should apply directly to the Hamiltonian restricted to the constraint
surface, which involves only
the physical degrees of freedom.  However, there is likely to be an analog
in generalized quantum dynamics of the methods used to treat constrained
quantum field theories (such as Faddeev-Popov determinants and
BRST invariance), which would permit working
in non--canonical gauges as well, and it would be interesting, and possibly
important, to find it.
Beyond these basically technical questions is of course the
larger issue of whether one can incorporate gravitation in a natural way,
and whether one can find a compellingly beautiful total trace dynamics which
gives rise to all of the observed forces and matter fields.

\bigskip
\centerline{\bf Acknowledgments}

This work was supported in part by the Department of Energy under
Grant \#DE--FG02--90ER40542.  One of the authors
(S. L. A.) wishes to acknowledge the
hospitality of the Aspen Center for Physics, and both the
Department of Applied
Mathematics and Theoretical Physics and Clare Hall at Cambridge University,
where parts of this work were done.
He also wishes to thank S. B. Treiman for suggesting the study of canonical
transformations, F. J. Dyson for expressing strong reservations
with approximations employed in the initial formulation of this work,
Y. Suhov for several important conversations about statistical mechanics, and
A. Kempf and G. Mangano for asking about corrections to the canonical algebra.
The other author (A. C. M.) wishes to thank T. Tao for valuable discussions.
\bigskip
\vfill\eject
\centerline{{\bf Appendix A.  Inclusion of Adjointness Restrictions in the}}
\centerline{{\bf Argument for Invariance of the Phase Space Measure}}
Inspection
of the argument of Eqs.~(39--41)
shows that the diagonal $(m=n)$ and off--diagonal
$(m \neq n)$ terms in the sum $\Sigma$ vanish separately, and for each of
these,
the summed contribution from the canonical coordinate and momentum
pair $q_r, p_r$ for each fixed $r$ also vanishes separately.  This observation
permits us to take the adjointness restrictions into account; in the
following discussion we shall write $d\mu = d\mu_B d\mu_F$, with $d\mu_B$
and $d\mu_F$ respectively the bosonic and fermionic integration measures.
There are three cases to be considered:\hfill \break
(1)  For a bosonic pair of phase space variables $q_r, p_r$,
the $x_r$ variables are independent but are both self--adjoint, and thus
$$(x_r)_{mn}^A= \epsilon^A (x_r)_{nm}^A~~~. \eqno(A1a) $$
This means that the integration
measure must be redefined to include all diagonal terms in $m,n$, but only
the upper diagonal off--diagonal terms, so that the bosonic integration
measure becomes
$$\eqalign{
d\mu_B=&\prod_A d \mu_B^A~~~, \cr
d\mu_B^A \equiv& \prod_{r,m \leq n} d(x_r)_{mn}^A~~~.\cr
}\eqno(A1b)$$
The argument for the diagonal terms in this product proceeds just as did that
for the diagonal terms in the unrestricted case, while the argument for the
off--diagonal terms uses Eq.~(A1a) in place of an interchange of the summation
index pair $m,n$, together with the fact that for a boson $\epsilon_r=1$,
again leading to the conclusion that the diagonal and off--diagonal
contributions to $\Sigma$ vanish independently.\hfill \break
(2)  For a fermionic pair of phase space variables in complex Hilbert space
constructed according to
the recipe $q_r=\psi_r,~~ p_r=i \psi_r^{\dagger} = iq_r^{\dagger}$,
the $x_r$ variables are
no longer independent.  However, this construction implies that
$$ (q_r)_{mn}^1= (p_r)_{nm}^0~,~~ (p_r)_{mn}^1=(q_r)_{nm}^0~~~, \eqno(A2)$$
and thus, in a complex Hilbert space, the fermionic integration measure
must be redefined as
$$\eqalign{
d\mu_F=& d \mu_F^0~~~, \cr
d\mu_F^0 \equiv& \prod_{r, m ,n}  d(x_r)_{mn}^0~~~.\cr
}\eqno(A3a)$$
Similarly, for the analogous fermionic construction $p_{r1}={1 \over 2}
\psi_r^{\dagger} I,...$ in a quaternionic Hilbert space, the fermionic
integration measure must be redefined as
$$\eqalign{
d\mu_F=&\prod_{A=0,2} d \mu_F^A~~~, \cr
d\mu_F^A \equiv& \prod_{r, m, n}  d(x_{r1})_{mn}^A~~~.\cr
}\eqno(A3b)$$
Since the argument for the unrestricted
case worked for each $A$ value separately, it still goes through as before.
\hfill \break
(3)  Finally, for a fermionic pair of phase space variables constructed
using a real representation of the imaginary unit,
with a pair of fermions for each $r$ obeying $q_{r2}=p_{r1}^{\dagger}~,~~
p_{r2}=-q_{r1}^{\dagger}$ but with no relation between $p_{r1}$ and $q_{r1}$,
one simply omits the variables $x_{r2}$ from
$d\mu_F$ and uses the unrestricted form of the fermionic measure for
the variables $x_{r1}$, and the invariance argument then proceeds just as
before.
\vfill \eject
\centerline{{\bf Appendix B.  Canonical Form of $\langle \tilde C
\rangle_{AV}$ and}}
\centerline{{\bf Implications for the Structure
of $x_r$ and $\tilde \lambda$}}

We give here the decomposition of the phase space operators $\{x_r\}$ with
respect to the canonical form for $\langle \tilde C \rangle_{AV}$
given in Eq.~(46a)
and the relationship this implies between $x_r$ and $x_{r~eff}$.   We also
discuss the canonical form for $\tilde \lambda$ which corresponds to that for
$\langle \tilde C \rangle_{AV}$  .

\parindent=0pt
(1) {\bf Real Hilbert space.}
In real Hilbert space, the anti--self--adjoint operator $\langle \tilde C
\rangle_{AV}$
is skew symmetric, and when the Hilbert space is even dimensional can be
brought by a real unitary ( i.e., orthogonal) transformation to the
canonical form $\langle \tilde C \rangle_{AV} = i_2 \otimes C_d$, with
$C_d$ a real diagonal matrix and with $i_2$ the $2 \times 2$
skew symmetric matrix
$$i_2=\left(\matrix{0&-1\cr1&0\cr}\right)~~~.\eqno(B1)$$
(When the Hilbert space is odd dimensional the canonical form consists of a
block of the form $i_2 \otimes C_d$,
and one further element $0$ on the principal
diagonal which does not correspond to a symplectic structure.)
The matrix $i_2$ spans a two dimensional real Hilbert
subspace, and a complete
set of operators [2] in this subspace can be taken
as $1_2,~i_2,~W$ and $W i_2$, with
$$
1_2=\left(\matrix{1&0\cr0&1\cr}\right)~,~~~
W=\left(\matrix{1&0\cr0&-1\cr}\right)~,~~~
Wi_2=\left(\matrix{0&-1\cr-1&0\cr}\right)~,~~~
\eqno(B2)$$
from which one sees that $i_2$ and $W$ anticommute.  The operator $x_r$
can now be expanded over the operator basis provided by Eqs.~(B1, B2), with
coefficients $x_{rA}~,~~A=0,1,2,3$ which are still operators
but which commute with $i_2$ and $W$,
$$x_r=x_{r0}1_2+x_{r1}i_2+x_{r2}W+x_{r3}Wi_2~~~.\eqno(B3a)$$
It is convenient to rewrite this expansion in terms of ``complex'' components
denoted by $x_{ra},~ x_{rb}$ which both commute with $i_2$, according to
$$\eqalign{
x_r=&x_{ra}+ Wx_{rb}~~~,\cr
x_{ra}=&x_{r0}1_2+x_{r1}i_2~,~~~x_{rb}=x_{r2}+x_{r3}i_2~~~.\cr
}\eqno(B3b)$$
Writing now $C_d={\cal E}D$, with $D$ nonnegative and with ${\cal E}$ a
diagonal matrix with elements $\pm 1$, and
writing $i_{eff}=i_2 {\cal E}$ (with a direct product $\otimes$ understood),
we find
$$x_{r~eff}={1 \over 2}[x_{ra}+Wx_{rb}-i_2 {\cal E} (x_{ra}+Wx_{rb}) i_2
{\cal E}]={1 \over 2}[x_{ra}+{\cal E} x_{ra} {\cal E} + W(x_{rb}-{\cal E}
x_{rb} {\cal E})]~~~,\eqno(B4a)$$
which reduces in the special case when ${\cal E}=1$ to
$$x_{r~eff}=x_{ra}~~~.\eqno(B4b)$$\hfill \break
\parindent=0pt
(2)  {\bf Complex Hilbert space.}
In complex Hilbert space an anti--self--adjoint operator can always be
written as the $c$--number $i$ times a self--adjoint operator,
and a self--adjoint operator
can always be brought by a complex unitary transformation to real diagonal
form.  So we have the canonical form $\langle \tilde C\rangle_{AV}=i C_d$;
writing $C_d={\cal E}D$, with $D$ nonnegative and with ${\cal E}$ again
a diagonal matrix with elements $\pm 1$, and writing
$i_{eff}=i {\cal E}$, we have
$$x_{r~eff}={1 \over 2}(x_r-i {\cal E} x_r i{\cal E})
={1 \over 2} (x_r + {\cal E} x_r {\cal E})~~~, \eqno(B5a)$$
which reduces in the special case when ${\cal E}=1$ to
$$x_{r~eff}=x_r~~~.\eqno(B5b)$$\hfill \break
\parindent=0pt
(3)  {\bf Quaternionic Hilbert space.}
In quaternionic Hilbert space the spectral analysis
for an anti--self--adjoint
operator differs in a nontrivial way from that for a self--adjoint operator
(see [2] for a detailed discussion and references), and implies that
by a quaternion unitary transformation, $\langle \tilde C \rangle _{AV}$
can be brought
to the form $I D$, with $I$ and $D$
commuting operators of the form
$$\eqalign{
I=&\sum_n |n\rangle i \langle n|~~~,\cr
D=&\sum_n |n \rangle D_n \langle n|~~~,\cr
}\eqno(B6a)$$
and with $D_n$ real and nonnegative.
We adjoin to this set of operators the additional two operators $J, K$, chosen
to commute with $D$ and to form a quaternion algebra with $I$; one
possible (but not unique) choice for these operators is
$$\eqalign{
J=&\sum_n |n\rangle j \langle n|~~~,\cr
K=&\sum_n |n\rangle k \langle n|~~~,\cr
}\eqno(B6b)$$
with $i,~j,~k$ quaternion scalars.  [The $J,~K$ of Eq.~(B6b) commute
with $(-1)^F$. In a Hilbert space with equal numbers of bosonic and
fermionic states, so that they can be put into one to one correspondence,
it is easy to construct an
alternative set of operators $J,~K$ which anticommute
with $(-1)^F$.]  Let us now expand the operator $x_r$ over the basis $1,~I,~J,
{}~K$ with formally real expansion coefficients $x_{rA}~,~~A=0,1,2,3$, which
commute with the entire $I~,J~,K$ quaternion algebra (the theory of this
is explained in detail in [2]), giving
$$x_r=x_{r0}+x_{r1}I+x_{r2}J+x_{r3}K~~~.\eqno(B7a)$$
It is convenient to rewrite Eq.~(B7a) in terms of formally complex
so--called symplectic components $x_{r\alpha},~x_{r\beta}$ which
both commute with $I$, according to
$$\eqalign{
x_r=&x_{r\alpha}+J x_{r\beta}~~~,\cr
x_{r\alpha}=&x_{r0}+x_{r1}I~,~~~x_{r\beta}=x_{r2}-x_{r3}I~~~.
}\eqno(B7b)$$
Writing now  $i_{eff} =I$, we get
$$x_{r~eff}={1 \over 2}[x_{r\alpha}+Jx_{r\beta}-I (x_{r\alpha}+Jx_{r\beta})
I]={1 \over2} [x_{r\alpha}(1+1)+Jx_{r\beta}(1-1)]=x_{r\alpha}~~~.\eqno(B8)$$
\parindent=25pt

Although a canonical form for $\tilde \lambda$ is not needed for the Ward
identity derivation of Sec.~6, it is nonetheless instructive to examine the
implications that the canonical form for $\langle \tilde C \rangle_{AV}$
has for
$\tilde \lambda$.  According to Eq.~(45b), there is a function $F$ for which
$$\langle \tilde C \rangle_{AV}=F(\tilde \lambda, (-1)^F)~,~~~
\eqno(B9a)$$
which implies the inverse functional relation
$$\eqalign{
\tilde \lambda=&G(\langle \tilde C \rangle_{AV}, (-1)^F)~,\cr
=&G_1(\langle \tilde C \rangle_{AV}, (-1)^F)+\langle \tilde C \rangle_{AV}
G_2(\langle \tilde C \rangle_{AV},(-1)^F)~,~~~\cr
}\eqno(B9b)$$
where $G_1$ and $G_2$ are even functions of the argument $\langle \tilde C
\rangle_{AV}$.  When we specialize to the form
$\langle \tilde C \rangle_{AV}=i_{eff} D$, with $D$ a real positive
$c$--number, as
assumed in Eq.~(46a), the
functions $G_{1,2}$ reduce to $c$--number functions of $(-1)^F$.  Moreover,
$G_1$ must be anti--self--adjoint, which implies that it must vanish in real
and quaternionic Hilbert spaces, giving in these cases the canonical form
$$\tilde \lambda =i_{eff} D [G_2^0+G_2^1(-1)^F]~,~~~\eqno(B10a)$$
with $G_2^{0,1}$ real constants.

In complex Hilbert space, $i$ is an
anti--self--adjoint $c$--number, so we get the canonical form
$$\tilde \lambda =i[G_1^0+G_1^1(-1)^F]+i_{eff} D [G_2^0+G_2^1(-1)^F]~,~~~
\eqno(B10b)$$
with $G_{1,2}^{0,1}$ real constants.  To achieve a further simplification
in this case, we assume that ${\bf {\hat H}}$ and ${\bf H}$ can be expressed
in terms of the phase space opertors $\{x_r\}$ using only real number
coefficients.  This implies that $\langle \tilde C \rangle_{AV}$, and hence
$i_{eff}$, can depend on $i$ only through $\tilde \lambda$, and conversely,
that $\tilde \lambda$ can depend on $i$ only through $i_{eff}$;
then the function
$G_1$ must vanish, and Eq.~(B10b) reduces to the simpler form given in
Eq.~(B10a).  More generally, this
implies that in complex Hilbert space,
ensemble averages of monomials constructed from
the phase space variables can depend on $i$ only through $i_{eff}$, a result
used at the end of Sec.~6.
\bigskip
\centerline{{\bf Appendix C.  Evaluation of the $\delta x_s$
Variations of $\tilde C$, ${\bf H}$, ${\bf \hat H}$, and $\tilde F$}}

Varying the definition
$$\tilde C=\sum_{r,s} x_r \omega_{rs} x_s  \eqno(C1a)$$
with respect to $x_s$, we get
$$\delta_{x_s} \tilde C=\sum_r( \delta x_s \omega_{sr} x_r
+x_r \omega_{rs} \delta x_s)~,~~~\eqno(C1b)$$
which by Eq.~(10c) becomes
$$\delta_{x_s} \tilde C=\sum_r \omega_{rs} (x_r \delta x_s -
\delta x_s \epsilon_r x_r)~~~.\eqno(C1c)$$

Varying ${\bf H}$ with respect to $x_s$ and using the definition of operator
derivative, we get
$$\delta_{x_s} {\bf H} ={\bf Tr} {\delta {\bf H} \over \delta x_s}
\delta x_s   ~~~.\eqno(C2a)$$
Applying Eq.~(10c) to Eq.~(9b)
then gives
$$\delta_{x_s} {\bf H} ={\bf Tr} \sum_r \dot x_r \omega_{rs}  \delta x_s
{}~~~.\eqno(C2b)$$

We turn our attention next to ${\bf \hat H}={\bf {\hat Tr}}H
={\rm Re}{\rm Tr} H$, which is naturally
conserved under the trace dynamics generated by ${\bf \hat L}
= {\bf{\hat Tr}}L={\rm Re}{\rm Tr} L$,
where $L$ is the same self--adjoint
operator Lagrangian as appears in the graded
total trace Lagrangian ${\bf L}={\bf Tr} L ={\rm Re Tr}(-1)^F L$.
Since we are assuming that ${\bf \hat H}$ is conserved under the equations of
motion for the graded trace dynamics, the Euler--Lagrange equations for
${\bf \hat L}$ and ${\bf L}$ must agree, up to an overall sign $\eta_r =\pm1$
which can be chosen independently for each degree of freedom,
$$\eqalign{
{\delta {\bf \hat L} \over \delta \dot q_r}=& \hat p_r = \eta_r
{\delta {\bf L} \over \delta \dot q_r} = \eta_r p_r~,~~~\cr
{\delta {\bf \hat L} \over \delta q_r}=&\eta _r {\delta {\bf L} \over \delta
q_r}
{}~.~~~\cr
}\eqno(C3)$$
In fact, for Lagrangians in which fermion time derivative terms all have
the structure $\psi_l^{\dagger}..... \dot \psi_r$, no cyclic
permutation is involved in varying with respect to $\dot \psi_r$ since this
already stands to the right, and so $\eta_r=1$ for all $r$
and $\hat p_r =p_r$, a result that will
be assumed henceforth.  Making a Legendre transformation
from ${\bf \hat L}$ to ${\bf \hat H}$,
$${\bf \hat H}={\bf \hat{Tr}} \sum_r p_r \dot q_r -{\bf \hat
L}~,~~~\eqno(C4a)$$
we find
$$\delta {\bf \hat H}={\bf \hat{Tr}} \sum_r(\dot q_r \delta p_r - \dot p_r
\delta q_r)~.~~~\eqno(C4b)$$
Thus the variation of ${\bf {\hat H}}$ is given by
$$\delta_{x_s}{\bf \hat H}= {\bf \hat{Tr}}\sum_r
( \dot q_r \delta_{x_s} p_r -\dot p_r \delta_{x_s} q_r)
={\bf Tr}(-1)^F \sum_r \dot x_r \hat \omega_{rs} \delta x_s~,~~~\eqno(C5a)$$
with $\hat \omega = {\rm diag}(\Omega_B,\Omega_B,...,
\Omega_B)$ in the notation of Eqs.~(10a, b).  Forming the sum which
is needed in the Ward identity derivation, we find
$$\eqalign{
\sum_s \omega_{us} \hat \omega_{rs} \equiv& \alpha_{ur}~~~,\cr
\alpha =&{\rm diag}(1_2,...,1_2,-W,...,-W)~~~,\cr
}\eqno(C5b)$$
with $1_2$ (for the bosonic variables) and $W$
(for the fermionic variables) as defined in Eq.~(B2).

Carrying out an analog of the discussion of Sec.~3, now using the ungraded
trace Hamiltonian ${\bf \hat H}$, we find a second conserved operator ${\hat
{\tilde C}}$ given by
$${\hat{\tilde C}} = \sum_{r,s} x_r \hat \omega_{rs} x_s
=\sum_{r,B} [q_r,p_r] + \sum_{r,F} [q_r,p_r]~~~. \eqno(C6a)$$
Defining the auxilliary conserved operator $\tilde F=-{1 \over 2}
(\tilde C - {\hat {\tilde C}})$, we have
$$\tilde F=\sum_{r,F} q_r p_r~~~,\eqno(C6b)$$
which when restricted to the canonical algebra is
(up to an additive constant) an anti--self--adjoint
version of the fermion number operator.   To take $\tilde F$ into account, we
include a term ${\bf Tr} \tilde \kappa \tilde F$,
or equivalently, a term ${\bf {\hat Tr}}
{\hat {\tilde \lambda} \hat{\tilde C}}$, in the exponent of
the equilibrium distribution;
the latter makes the algebraic calculations more symmetric and is thus more
convenient.  We now make the essential
assumption that the ensemble averages of $\tilde F$
and $\hat {\tilde C}$ are functions solely of the ensemble average of
$\tilde C$ and of the grading operator $(-1)^F$,
or equivalently, that the anti--self--adjoint
operator ensemble parameters $\tilde \kappa$ and ${\hat {\tilde \lambda}}$
are functions solely of $\tilde \lambda$ and of $(-1)^F$,
and thus commute with $\tilde \lambda$.  The validity of this assumption
is demonstrated in Appendix G, where we study implications of the full
Ward identity structure.  A
straightforward calculation shows that
$$\delta_{x_s}  {\bf {\hat Tr}} {\hat {\tilde \lambda} \hat{\tilde C}}
={\bf Tr} (-1)^F [\hat {\tilde \lambda}, \sum_r \hat \omega_{rs} x_r]
\delta x_s~~~, \eqno(C7a)$$
which has a similar commutator structure to the relation
$$\delta_{x_s} {\bf Tr} \tilde \lambda \tilde C=
{\bf Tr} [\tilde \lambda, \sum_r \omega_{rs} x_r] \delta x_s~~~\eqno(C7b)$$
deduced from Eq.~(C1c).  Consequently, the commutativity of the ensemble
parameters allows us to use the argument of Eqs.~(54a--c) to conclude that
the variation of the $\hat {\tilde C}$ term in the equilibrium distribution
does not contribute to the Ward identity.  The functional relation assumption
also implies the continuing validity of the other parts of our
analysis that depended on the
structure of $\tilde \lambda$.  Another interesting consequence of
commutativity of the ensemble
parameters is that it implies a vanishing generalized Poisson
bracket of the $\tilde C$ term in the equilibrium distribution with the
$\hat {\tilde C}$ term, with the result that the four first integrals
appearing in the equilibrium distribution all
have vanishing generalized Poisson
brackets with one another.  Inclusion of the $\hat {\tilde C}$ term in the
equilibrium distribution of course also implies that this term now appears,
in a role analogous to that of the $\tilde C$ term, in the thermodynamic
expressions of Eqs.~(49a--d).
\bigskip
\centerline{{\bf Appendix D.  Proof of the Lemma of Sec. 6}}

We restate and then prove the Lemma used in Sec.~6 to take account of
adjointness restrictions on the variations.

\leftline{\bf Lemma:}  Let $Y_1$ and $Y_2$ be two self--adjoint bosonic
or two anti--self--adjoint bosonic operators constructed
from the phase space variables.  Then in $0=\delta
{\bf Tr}Y_1 Y_2$, the self--adjointness restrictions on the variations can
be ignored.\hfill \break

{\leftline{\bf Proof:}}  By the cyclic property of {\bf Tr}, we have
$$\eqalign{
{\bf Tr} Y_1 Y_2 =& {\bf Tr} Y~~~,\cr
Y\equiv& {1 \over 2} (Y_1Y_2+Y_2Y_1)=Y^{\dagger}~~~,\cr
}\eqno(D1)$$
and so it suffices to prove that in $0=\delta {\bf Tr}Y $ with manifestly
self--adjoint
$Y$, the self--adjointness restrictions on the variations can be ignored.

We consider first the case of the variation of a bosonic variable $x_r$, for
which the self--adjointness of $x_r$ implies that
$(\delta x_r)^{\dagger}=\delta x_r$. Self--adjointness of $Y$ implies
that for each term in $Y$ of the form
${\cal O}_L x_r {\cal O}_R$ there must be a corresponding
term ${\cal O}_R^{\dagger} x_r {\cal O}_L^{\dagger}$, with the grade
$\epsilon_L$ of ${\cal O}_L$ equal to the grade $\epsilon_R$ of ${\cal O}_R$
for there to be a nonvanishing graded trace.  The summed contribution of the
two terms when $x_r$ is varied is
$${\bf Tr} \epsilon_R ({\cal O}_R {\cal O}_L
+ {\cal O}_L^{\dagger} {\cal O}_R^{\dagger}) \delta x_r~~~,\eqno(D2)$$
and since the coefficient of $\delta x_r$ is manifestly
self--adjoint we are justified
in equating the coefficient of $\delta x_r$ to zero.

We consider next the
variation of a fermionic variable $x_r$, for which there is another
fermionic variable $x_{s(r)}$ for which $x_r^{\dagger}=c_rx_{s(r)}$ and
$x_{s(r)}^{\dagger}= c_r x_r$, with $c_r$ a $c$--number of unit magnitude
with conjugate $\bar c_r$, so that $c_r \bar c_r=\bar c_r c_r=1$.
(The methods for including fermions described in Sec.~3 take this form
with either
$c_r=-i$ or  $c_r=\pm 1$.)  The corresponding
variations must thus be related by the self--adjointness restriction
$\delta x_r^{\dagger}=c_r \delta x_{s(r)}.$
Self--adjointness of $Y$ now
implies that for each term in $Y$ of the form
${\cal O}_L x_r {\cal O}_R$ there must be a corresponding
term ${\cal O}_R^{\dagger}c_r x_{s(r)} {\cal O}_L^{\dagger}$, with the grade
$\epsilon_L$ of ${\cal O}_L$ opposite to the
grade $\epsilon_R$ of ${\cal O}_R$
for there to be a nonvanishing graded trace.  The summed contribution of the
two terms when $x_r$ is varied is then
$${\bf Tr}( \epsilon_R {\cal O}_R {\cal O}_L \delta x_r
+ \epsilon_L {\cal O}_L^{\dagger} {\cal O}_R^{\dagger}
c_r \delta x_{s(r)})~~~.\eqno(D3)$$
The self--adjointness restriction on the variations implies that the second
term in Eq.~(D3) is equal to
$${\bf Tr} \epsilon_L {\cal O}_L^{\dagger} {\cal O}_R^{\dagger}
\delta x_r^{\dagger} ~~~,\eqno(D4)$$
which using the fact that {\bf Tr} of any operator is equal to {\bf Tr} of
the adjoint of the same operator, is equal to
$${\bf Tr} \epsilon_L \delta x_r {\cal O}_R {\cal O}_L
={\bf Tr} (-\epsilon_L) {\cal O}_R {\cal O}_L \delta x_r
={\bf Tr} \epsilon_R {\cal O}_R {\cal O}_L \delta x_r~~~,\eqno(D5)$$
which just doubles the contribution from the first term in Eq.~(D3).
Hence we get the correct answer by equating the coefficients of $\delta x_r$
and $\delta x_{s(r)}$ independently to zero in Eq.~(D3).  \hfill\break
\bigskip
\centerline{{\bf Appendix E. Use of the Sources to Generate the Polynomial }}
\centerline{{\bf $P(\{ x_r \})$ in Eqs.~(57d) and (61e)}}

We show here that by variation of the sources in Eqs.~(57c) and (61c), one
obtains Eqs.~(57d) and (61e), in which ${\partial \over \partial t}$,
$\delta$, $[\tilde H_{eff},~~]$, and $[\tilde G_{eff},~~]$ all act on the
Weyl ordered
polynomial $P(\{ x_r \})$ by the Leibnitz product rule.  Our argument also
applies to Eq.~(60b) (after dropping the $\hat \tau$ and $\tau$ terms), and
shows that when $V$ in Eq.~(60d) is Weyl ordered,
the commutator appearing in Eq.~(60d) can be
evaluated in terms of the canonical algebra of Eq.~(13) by the
Leibnitz product rule.  Representing
Eq.~(57c), Eq.~(60b) (with the parentheses indicating implicit summation
restored and $\sigma_t$ replaced by $\sigma^{\prime}_t$, so that
the terms inside the expectation read
$[i_{eff} (\sigma^{\prime}_t x_{t~eff}), x_u]
-\hbar \omega_{ut} \sigma^{\prime}_t~$), and Eq.~(61c),
after multiplication through by the partition
function $Z$, by the generic structure
$$0=Z\langle {\cal D}x_u \rangle_{AV}~~~,\eqno(E1a)$$
we wish to show that by varying the sources in Eq.~(E1a) we can also derive
$$0=Z\langle {\cal D} P(\{ x_r \}) \rangle_{AV}~~~,\eqno(E1b)$$
where ${\cal D}$ acts on the Weyl ordered polynomial
$P$ by the Leibnitz product rule
$${\cal D}(x_rx_s)=({\cal D}x_r) x_s + x_r({\cal D}x_s)~~~.\eqno(E1c)$$

We begin by observing that Eq.~(E1a) also implies that
$$0=Z\langle {\cal D}(\sigma_u x_u) \rangle_{AV}~~~,\eqno(E2a)$$
and we shall work with the self--adjoint bosonic variables $(\sigma_u x_u)$
henceforth.  Multiplying Eq.~(E2a) by ${\bf Tr} \delta \rho_u$, with
$\delta \rho_u$ an arbitrary self--adjoint bosonic
operator, Eq.~(E2a) becomes
$$0=Z\langle {\bf Tr} \delta \rho_u {\cal D}(\sigma_u x_u) \rangle_{AV}~~~,
\eqno(E2b)$$
which we take as the starting point for our discussion.
Writing $P(\{ x_r \})=S[\{ (\sigma_r x_r) \}]$, with $S$ a
totally symmetrized polynomial in its arguments, application of Eq.~(E1c)
gives
$${\cal D} S[\{ (\sigma_r x_r) \}]=\sum_u S[\{ (\sigma_r x_r), r\not=u\};
{\cal D} (\sigma_u x_u)]~~~,\eqno(E2b)$$
where the sum over $u$ ranges over the indices of
all variables $x_r$ which appear as
arguments of $P$.  Hence to derive Eq.~(E1b) it suffices to derive
$$0=Z\langle \sum_u S[\{ (\sigma_r x_r), r\not=u\}; {\cal D} (\sigma_u x_u)]
\rangle_{AV}~~~.\eqno(E3)$$

We begin by varying Eq.~(E2b) with respect to the
source corresponding to each
$r \not =u$ (if a variable $x_R$ appears multiple times, we perform
multiple independent variations with respect to its source $\rho_R$),
and after taking these variations, then
summing over all choices of $u$ from the among the indices appearing in $P$.
There are two types of source dependence which contribute:  there are the
source terms in the equilibrium distribution of Eq.~(51a)
that we use to form the ensemble averages, and also the
explicit source term in the Ward identity that was suppressed in
writing Eqs.~(57c), (60b), (61c), and (E2b).   From Eq.~(55b) we see that the
contribution to the summand of
this latter term, after left multiplication
by ${\bf Tr} \delta \rho_u \sigma_u$, is equal to
$$Z\langle {\bf Tr} \delta \rho_u\sum_s \omega_{us} \sigma_u \sigma_s \rho_s
\hbar {\bf V}\rangle_{AV}~~~,\eqno(E4a) $$
with ${\bf V}$ appropriate to one of the applications discussed in the text.
When Eq.~(E4a) is varied
with respect to the source $\rho_v$, it contributes
$$Z\langle {\bf Tr} \delta \rho_u \omega_{uv} \sigma_u \sigma_v \delta \rho_v
\hbar {\bf V}\rangle_{AV}~~~,\eqno(E4b) $$
which can be rewritten as
$$Z\langle {\bf Tr} \delta \rho_u \delta \rho_v
\hbar {\bf V}\rangle_{AV} \omega_{uv} \sigma_u \sigma_v ~~~.\eqno(E4c) $$
Now for each term with the form of Eq.~(E4c) in the sum over $u$,
symmetrization implies that there is a corresponding term with the roles of
$u$ and $v$ interchanged, giving a total contribution of
$$Z\langle {\bf Tr} \delta \rho_u \delta \rho_v \hbar {\bf V}\rangle_{AV}
(\omega_{uv} \sigma_u \sigma_v+\omega_{vu} \sigma_v \sigma_u)
{}~~~.\eqno(E4c) $$
But this vanishes because for bosonic $u,v$, the auxilliary quantities
$\sigma_u$ and $\sigma_v$
commute and $\omega_{uv}+\omega_{vu}=0$, while for fermionic $u,v$, we have
$\omega_{uv}=\omega_{vu}$ and  $\{\sigma_u,\sigma_v\}=0$.  Hence the explicit
source term in the Ward identity makes no contribution to symmetrized
expressions.  In the remaining terms in the
sum over $u$, the variations $\delta \rho_u$ and $\delta \rho_v$ each
appear in a separate graded trace.
Thus, after
implementing the cancellation of Eq.~(E4c), it suffices to show that we
can derive the generic term in the summand of Eq.~(E3) by operations on
a product of source variation factors, since once the generic term has the
correct symmetrized polynomial form, all terms in the sum over $u$ are
guaranteed to have this form.

At this point let us take advantage of the fact that the $\delta \rho$ are all
arbitrary self--adjoint bosonic operators, permitting
us to replace them by $(-1)^F \delta
\rho$, with the $\delta \rho$ again arbitrary self--adjoint
bosonic operators, thereby
converting all graded traces involving the source variations to ungraded
traces.
We are thus left with the simpler problem of deriving
$$0=Z\langle S[\{ (\sigma_r x_r), r\not=u\};
{\cal D} (\sigma_u x_u)]\rangle_{AV}~~~,\eqno(E5a)$$
given the identity
$$0=Z\langle \prod_{r \not =u} [{\rm Re Tr} (\sigma_r x_r) \delta \rho_r]
{\rm Re Tr} \delta \rho_u {\cal D} (\sigma_u x_u)\rangle_{AV}~~~.\eqno(E5b)$$
To do this, we exploit the arbitrariness of the variations $\delta \rho$,
as follows.  Let $\Lambda_A$ be a complete basis of trace normalized
bosonic self--adjoint operators, that is, this basis is characterized by the
properties that
$${\rm Tr} \Lambda_A \Lambda_B=\delta_{AB}~~~,\eqno (E6a)$$
and that any bosonic self--adjoint operator ${\cal O}_1$ can be expanded
in the form
$${\cal O}_1=\sum_A {\cal O}_{1A} \Lambda_A~,~~~
{\cal O}_{1A}={\rm Re Tr} {\cal O}_1 \Lambda_A~~~,\eqno(E6b)$$
which implies the formula
$${\rm Re Tr} {\cal O}_1 {\cal O}_2 =\sum_A {\rm Re Tr} {\cal O}_1 \Lambda_A
{\rm Re Tr} {\cal O}_2 \Lambda_A~~~.\eqno(E6c)$$
Now let us take $\delta \rho_R=\{\Lambda_A, \kappa \}$  and
$\delta \rho_S=\Lambda_A$ in Eq.~(E5b),
with $\kappa$ an arbitrary bosonic self--adjoint operator,
and sum over $A$.  By Eq.~(E6c) this leads to the replacement of the
product of factors ${\rm Re Tr} {\cal O}_R \delta \rho_R
{\rm Re Tr} {\cal O}_S
\delta \rho_S$ with the single factor
$${\rm Re Tr} \{ {\cal O}_R, \kappa \} {\cal O}_S
={\rm Re Tr}\kappa \{ {\cal O}_R, {\cal O}_S \}~~~,\eqno(E7a)$$
which involves the symmetrized (or Jordan) product of the two operators
${\cal O}_{R,S}$.  Proceeding in this fashion, and using the freedom of
the $\kappa$ operators just as we use the freedom of the operators $\delta
\rho$, we can build up from Eq.~(E5b) any trace identity of the form
$$0={\rm Re Tr} \kappa Z \langle \hat S[\{ (\sigma_r x_r), r\not=u\};
{\cal D} (\sigma_u x_u)]\rangle_{AV}~~~,\eqno(E7b) $$
with $\kappa$ an arbitrary bosonic self--adjoint operator,
which further implies the operator identity
$$0=Z\langle \hat S[\{ (\sigma_r x_r), r\not=u\};
{\cal D} (\sigma_u x_u)]\rangle_{AV}~~~,\eqno(E7c) $$
in which $\hat S$ is any self--adjoint polynomial
that can be constructed from
its arguments by repeated applications of the symmetrized product.  But
repeated application of the identity
$$2S(x_1,....,x_L)=\sum_w \{x_w, S(x_1,...,(x_w),...,x_L) \} ~~~,\eqno(E8)$$
with $(x_w)$ indicating that $x_w$ is to be omitted from the argument list,
shows that any totally symmetrized polynomial $S$ can be built up by repeated
applications of the symmetrized product to its arguments; hence $\hat S$ in
Eq.~(E7c) can be taken to be $S$, completing the
derivation of Eq.~(E5a)

Although we have phrased this derivation entirely in terms of symmetrizing
operations, it is likely that it can be significantly extended as follows.
If we take
$\delta \rho_R=[\Lambda_A, \tilde \kappa ]$, with $\tilde \kappa $ now
anti--self--adjoint, then Eq.~(E7a) is replaced by
$${\rm Re Tr} [ \tilde \kappa, {\cal O}_R ] {\cal O}_S
={\rm Re Tr}\tilde \kappa [{\cal O}_R, {\cal O}_S ]~~~,\eqno(E9)$$
and thus from the source variations we can in fact build up
polynomials which are antisymmetrized in some variables.
Moreover, the argument for the vanishing of the explicit
source term contribution requires not total symmetrization, but only
symmetrization in all aguments $x_u,~x_v$ for which the symplectic structure
$\omega_{uv}$ is nonzero.  Thus, if we define a {\it partially Weyl ordered}
polynomial to be a polynomial which is symmetrized with respect to all
arguments $x_u,~x_v$ for which $\omega_{uv} \not=0$, then it appears likely
that with careful attention to the combinatorics, one should be able to show
that Eq.~(E1a) implies the extension of Eq.~(E1b) in which $P$ is any
partially Weyl ordered polynomial in its arguments.   The example of
Eq.~(17e) also indicates that there will be extensions of Eq.~(E1b) to
certain cases in which $P$ is a non--Weyl ordered polynomial of low degree.
We expect these cases to play an important role in the study of operator
gauge invariant theories, which we will take up in detail elsewhere. \hfill
\break
\bigskip
\centerline{{\bf Appendix F. Canonical and Symmetry Transformations}}

We give here further details of the structure of canonical transformations in
generalized quantum dynamics, with special emphasis on their role as symmetry
transformations.

A particularly interesting class of canonical
transformations are what we shall term linear symmetry transformations,
defined as transformations of the type Eq.~(27) generated by
total trace functionals of the special form
$$\eqalign{
{\bf G}_h=&{\bf Tr} G_h~~~, \cr
G_h =&\sum_{r,s} p_r h_{rs} q_s~~~. \cr
}\eqno(F1a)$$
These transformations linearly transform the canonical
coordinates $\{ q_r \}$ among
themselves, with a corresponding transformation on the canonical momenta,
but do not mix coordinates with momenta.  When the indices $r,s$ in Eq.~(F1a)
are both bosonic or both fermionic, the coefficients $h_{rs}$ are taken to be
ordinary $c$--numbers, while when one index is fermionic and one is
bosonic, the coefficients $h_{rs}$ are taken to be
Grassmann $c$--numbers. Thus, the linear symmetry
transformations include grade--changing transformations which mix the bosonic
and fermionic coordinates.  As a consequence of the grading structure of
$h$, we have
$${\bf Tr} p_r h_{rs} q_s = {\bf Tr} (\pm) h_{rs} p_r q_s~~~,\eqno(F1b)$$
which together with the cyclic property of {\bf Tr} implies that $G_h$
is Weyl ordered.  Thus, a
linear symmetry transformation is also a Weyl ordered intrinsic
canonical transformation. Under the generalized Poisson bracket operation,
two linear symmetry transformations compose as [1, 2]
$$\{ {\bf G}_g, {\bf G}_h \} ={\bf G}_{[g,h]} ~~~,\eqno(F2a)$$
with $[g,h]$ the commutator
$$[g,h]_{rs} = \sum_t (g_{rt} h_{ts} - h_{rt} g_{ts})~~~,\eqno(F2b)$$
and hence
linear symmetry transformations form a Lie commutator
algebra under the generalized
Poisson bracket.  Clearly a linear
rearrangement of the canonical coordinates among
themselves, with $c$--number or Grassmann $c$--number coefficients,
together with a corresponding linear transformation among the momenta,
transforms a Weyl ordered polynomial into another such polynomial.
Therefore the set of Weyl ordered total trace functionals, and thus of
Weyl ordered intrinsic canonical transformations,
is closed under the action of
linear symmetry transformations. [We note in passing that in Refs.~[1, 2]
we also
introduced linear symmetry transformations in which $h$ is an arbitrary
quaternionic (hence non--commutative) coefficient matrix; although the Lie
property of Eqs.~(F2a, b) holds for this generalization, most of the other
properties of canonical transformations derived in Sec.~4 and
this Appendix do not.  For example, symmetry transformations based on
quaternionic representions of
compact Lie groups do not leave $\tilde C$ invariant.]
One can also define a generalization of linear symmetry
transformations with generators which can mix coordinates and momenta
according to
$$\eqalign{
{\bf G}_{\hat h}=&{\bf Tr} G_{\hat h}~~~, \cr
G_{\hat h}=&\sum_{r,s}x_r \hat h_{rs} x_s~~~,\cr
}\eqno(F2c)$$
with the grading structure of $\hat h_{rs}$ analogous to that for $h_{rs}$,
which implies that these transformations are also Weyl ordered.
These transformations also form a Lie algebra under the generalized Poissson
bracket, with the structure
$$\eqalign{
\{ {\bf G}_{\hat g} , {\bf G}_{\hat h} \}=&{\bf G}_{\hat k}~~~, \cr
\hat k_{tu} =&\sum_{r,s} (\hat g_{tr}+\epsilon_r \hat g_{rt}) \omega_{rs}
(\hat h_{us} +\epsilon_s \hat h_{su})  ~~~;\cr
}\eqno(F2d)$$
certain Bogoliubov transformations are of this more general type.

We shall now derive a second relation which is similar in structure to
Eq.~(29) of the text, and which describes the action of a linear symmetry
generator ${\bf G}_h = {\bf Tr} G_h$ on a Weyl ordered intrinsic canonical
generator ${\bf G}={\bf Tr} G$, when the phase space operators $\{ x_r \}$
are specialized to the canonical algebra,
$$\{ {\bf G}_h , {\bf G} \} =-{\bf Tr}i[G_h,G] ~~~.\eqno(F3)$$
Since by the Weyl ordering hypothesis $G$ is symmetrized,
we can represent it
as a sum of monomial terms produced by generating functions with the form
$g^n$ of Eq.~(11a), and it then suffices to prove the identity for only one
such term.  Writing $g$ in the form
$$g=\sum_r (\xi_r q_r + \eta_r p_r)~~~, \eqno(F4)$$
use of the generalized Poisson bracket in the form given in Eq.~(6b) gives
$$\{ {\bf G}_h, {\bf Tr} g^n \}=n {\bf Tr} g_h g^{n-1}~~~, \eqno(F5a)$$
with $g_h$ defined by
$$g_h=\sum_{r,s} (p_s \epsilon_r h_{sr} \eta_r -\xi_r h_{rs} q_s)~~~.
\eqno(F5b)$$
One can now check that over the canonical algebra
one has
$$ig_h=[\sum_{r,s}p_r h_{rs} q_s, \sum_t(\xi_t q_t +\eta_t p_t)]~~~.
\eqno(F5c)$$
But the commutator $[-iG_h, g^n]$ reduces to the totally symmetrized
product of $g_h$
with $g^{n-1}$, which is equal to the right hand
side of Eq.~(F5a) under the trace, completing the proof of Eq.~(F3).
Thus, there is an isomorphism between (a) the action of a
linear symmetry transformation on an arbitrary Weyl ordered intrinsic
canonical transformation, under the generalized Poisson
bracket operation of generalized quantum dynamics, and (b) the corresponding
behavior of the canonical algebra specializations of these transformations,
under the usual commutator operation. This isomorphism extends to
the more general Bogoliubov type transformation of Eq.~(F2c), where we
find
$$\eqalign{
\{ {\bf G}_{\hat h} , {\bf G} \} =&-{\bf Tr}i[G_{\hat h},G]
=n{\bf Tr} g_{\hat h} g^{n-1}~~~,\cr
g_{\hat h}=&\sum_{r,s,t} (\omega_{rt} \sigma_t \hat h_{rs} x_s
+ x_r \hat h_{rs} \omega_{st} \sigma_t) ~~~.\cr
}\eqno(F6)$$

The isomorphism does not extend, however, to the action of generic
Weyl ordered intrinsic canonical transformations on one another.
To see this, let us
consider the case of two such canonical transformations with generators
${\bf G}_1$ and ${\bf G}_2$ which are generating functions for Weyl ordered
monomials,
$$\eqalign{
&{\bf G}_1={\bf Tr} G_1,~~~ {\bf G}_2= {\bf Tr} G_2, \cr
&G_1=g_1^{n_1},~~~G_2=g_2^{n_2}, \cr
&g_{1,2}=\sum_r \sigma_{1,2r} x_r~~~.\cr
}\eqno(F7a)$$
Then from Eq.~(9a) we find
$$\eqalign{
\{ {\bf G}_1, {\bf G}_2 \} =& {\bf Tr} \sum_{r,s} n_1 g_1^{n_1-1}
\sigma_{1r} \omega_{rs} n_2 g_2^{n_2-1} \sigma_{2s} \cr
=& C {\bf Tr} g_1^{n_1-1} g_2^{n_2-1}~~~, \cr
C=&n_1 n_2 \sum_{r,s} \sigma_{1r} \omega_{rs} \sigma_{2s}~~~, \cr
}\eqno(F7b)$$
which is clearly not Weyl ordered when
$n_1$ and $n_2$ are both greater than 2. Thus the generalized Poisson
bracket of the generators for two Weyl ordered intrinsic
canonical transformations is in
general not a Weyl ordered canonical transformation.  On the other hand,
if we specialize to the canonical algebra and then evaluate the commutator
of $G_1$ and $G_2$, we easily find using Eq.~(13) that
$$[G_1,G_2] = iC S(g_1^{n_1-1}, g_2^{n_2-1})~~~,\eqno(F7c)$$
with $S$ the polynomial formed from completely symmetrizing $n_1-1$
factors of $g_1$ with respect to $n_2-1$ factors of $g_2$, which is Weyl
ordered.  In other words, over the canonical algebra, the commutator of
two Weyl ordered generators is Weyl ordered, but this does not correspond to
the composition properties of Weyl ordered canonical generators under the
generalized Poisson bracket operation.

In future work we plan to give a more detailed analysis of the Poincar\'e
transformations in generalized quantum dynamics than was given in [1, 2],
including a study of their relationship to the classification of canonical
transformations given in Sec.~4 and here.  \hfill\break
\bigskip
\centerline{{\bf Appendix G. The Full Ward Identity Structure}}
We describe here the full Ward identity structure resulting when the
presence of the additional conserved anti--self--adjoint operator
$\tilde F$ of Eq.~(C6b), or equivalently $\hat {\tilde C}$ of Eq.~(C6a),
is taken into account.
Including the latter in the equilibrium distribution, Eq.~(51a) becomes
$$\eqalign{
\rho=&Z^{-1} \exp[-{\bf Tr} \sum_r \rho_r (\sigma_r x_r)]
\exp(-{\bf {\hat Tr}} \hat {\tilde \lambda} \hat {\tilde C}
-{\bf Tr} \tilde \lambda \tilde C  - \hat{\tau} {\bf \hat H}-\tau
{\bf H})~~~, \cr
Z=&\int d\mu \exp[-{\bf Tr} \sum_r \rho_r (\sigma_r x_r) ]
\exp(-{\bf {\hat Tr}} \hat {\tilde \lambda} \hat {\tilde C}
-{\bf Tr} \tilde \lambda \tilde C-\hat{\tau} {\bf \hat H}-\tau
{\bf H})~~~; \cr
}\eqno(G1)$$
ensemble averages denoted by
the subscript ``$AV$'' will be understood henceforth to be taken in
the distribution of Eq.~(G1).  Including the $\hat {\tilde C}$ term in the
Ward identity of Eq.~(55b), and for the moment retaining {\it all} terms
coming from the variation of the equilibrium distribution, we get
$$\eqalign{
0=&\langle
([\tilde \lambda, x_u]
+(-1)^F[\hat{\tilde \lambda}, \sum_r \alpha_{ur} x_r]
+\hat{\tau}(-1)^F\sum_{r} \alpha_{ur} \dot x_r
+\tau \dot x_u +\sum_s \omega_{us} \sigma_s \rho_s ) \hbar {\bf V} \cr
+& [{1 \over 2} \{i_{eff}, V \}, x_u]
- \hbar \sum_s \omega_{us} {\delta {\bf V} \over \delta x_s}
\rangle_{AV}~,~~~ \cr
}\eqno(G2)$$
giving the full version of the Ward identity derived in Sec.~6.  Let us now
take $V=H$, and make the approximation (expected to be valid in the large
N limit) of replacing ${\bf H}$ by its ensemble epectation $\langle {\bf H}
\rangle_{AV}$.  The coefficient of this term in Eq.~(G2) then becomes the
proportional to the variation $\sum_s \omega_{us}
\delta Z / \delta x_s$ of the partition
function of Eq.~(G1), which is zero, and so Eq.~(G2) reduces, after use of
the equation of motion of Eq.~(9b) and division by $\hbar$, to
$$0=\langle
 [{1 \over 2} \{ \hbar^{-1} i_{eff}, H \}, x_u]
- \dot x_u
\rangle_{AV}~.~~~
\eqno(G3)$$
This is the same relation as was obtained in Eq.~(56b) of the text, but we
have now shown that its derivation does not depend on the assumption that the
ensemble parameters $\hat {\tilde \lambda}$ and $\tilde \lambda$ are
functionally related.

Let us now derive a second Ward identity by using properties of the
operator $\hat {\tilde C}$.  Since this is an anti--self--adjoint operator,
its ensemble expectation can be written in a polar form analogous to
Eq.~(46a) for $\langle \tilde C \rangle_{AV}$,
$$\eqalign{
&\langle \hat {\tilde C}\rangle_{AV}=\hat i_{eff} \hat D~,
{}~~~  \hat i_{eff}=-\hat i_{eff}^{\dagger}~,~~~
\hat i_{eff}^2=-1~,~~~\cr
&[\hat i_{eff},\hat D]=0~,~~~ \hat D~~{\rm self-adjoint~and~nonnegative}.~~~\cr
}\eqno(G4)$$
The difference is that we do not now assume $\hat D$ to be diagonal, but
we shall assume it to be nonsingular, so that the inverse $(\hat D)^{-1}$
exists.  Let us now derive a second Ward identity, by considering the
expression
$$0=\int d\mu \delta_{x_s}\left[ \rho {\bf \hat {Tr}} \{ \hat {\tilde C},
(\hat D)^{-1} \hat i_{eff} \} V \right]~~~, \eqno(G5a)$$
with $\rho$ the full equilibrium distribution of Eq.~(G1). Proceeding as
in Eqs.~(53--55) of the text, with the one exception that we now multiply
through by ${1 \over 2} \sum_s \hat \omega_{us}$, we end up with the following
analog of Eq.~(G2),
$$\eqalign{
0=&\langle
([\hat {\tilde \lambda}, x_u]
+(-1)^F[\tilde \lambda, \sum_r \alpha_{ur} x_r]
+\tau(-1)^F\sum_{r} \alpha_{ur} \dot x_r
+\hat{\tau} \dot x_u +(-1)^F\sum_s \hat \omega_{us} \sigma_s \rho_s )
 {\bf \hat V} \cr
+& [{1 \over 2} \{(\hat D)^{-1} \hat i_{eff}, V \}, x_u]
- \sum_s \hat \omega_{us} {\hat \delta {\hat {\bf V}} \over \hat \delta x_s}
\rangle_{AV}~,~~~ \cr
}\eqno(G5b)$$
where the hatted operator derivative
$\hat \delta {\hat {\bf V}} / \hat \delta x_s$
is defined by
$$ \delta {\hat {\bf V}}= {\bf {\hat Tr}} \sum_s
{\hat \delta {\hat {\bf V}} \over \hat \delta x_s} \delta x_s~~~.\eqno(G5c)$$

Let us now take $V=H$ in Eq.~(G5b), and replace the conserved extensive
quantity ${\bf \hat H}$ by its ensemble average $\langle {\bf \hat H}
\rangle_{AV}$.  The coefficient multiplying this quantity in Eq.~(G5b) now
becomes proportional to the partition function variation $\sum_s \hat
\omega_{us} \hat \delta Z/\hat \delta x_s$, which again vanishes, and so
Eq.~(G5c) reduces in this special case to
$$
0=\langle
 [{1 \over 2} \{(\hat D)^{-1} \hat i_{eff}, H \}, x_u]
- \sum_s \hat \omega_{us} {\hat \delta {\hat {\bf H}} \over \hat \delta x_s}
\rangle_{AV}~.~~~
\eqno(G6a)$$
But comparing the definition of Eq.~(G5c) with Eq.~(C4b), we see that the
fact that the same equations of motion follow from the ungraded and graded
trace variational principles can be expressed succinctly by the identity
$$\dot x_r =\sum_s \omega_{rs} {\delta {\bf H} \over \delta x_s}
=\sum_s \hat \omega_{rs} {\hat \delta {\bf \hat H} \over \hat \delta x_s}
{}~~~,\eqno(G6b)$$
and so Eq.~(G6a) simplifies further to
$$
0=\langle
 [{1 \over 2} \{(\hat D)^{-1} \hat i_{eff}, H \}, x_u]
- \dot x_u
\rangle_{AV}~.~~~
\eqno(G7a)$$
Comparing this expression for the ensemble average
of $\dot x_u$ (still in the presence of sources!)
with the similar expression obtained from the
original Ward identity in Eq.~(G3), we conclude that we must have the
relation
$$\hbar^{-1} i_{eff} = (\hat D)^{-1} \hat i_{eff}~~~,\eqno(G7b)$$
which since $\hat D$ is nonnegative implies the further relations
$$\eqalign{
&\hat D=\hbar~,~~~\hat i_{eff}= i_{eff}~,~~~  \cr
&\langle \hat {\tilde C} \rangle_{AV}=\langle \tilde C \rangle_{AV}~,~~~
\langle \tilde F \rangle_{AV}=0~,~~~\cr
}\eqno(G7c)$$
justifying the functional relation assumption made in Appendix C.
Substituting Eq.~(G7b) into the second Ward identity of Eq.~(G5b), and
multiplying through by $\hbar$, we get
$$\eqalign{
0=&\langle
([\hat {\tilde \lambda}, x_u]
+(-1)^F[\tilde \lambda, \sum_r \alpha_{ur} x_r]
+\tau(-1)^F\sum_{r} \alpha_{ur} \dot x_r
+\hat{\tau} \dot x_u +(-1)^F\sum_s \hat \omega_{us} \sigma_s \rho_s )
\hbar {\bf \hat V} \cr
+& [{1 \over 2} \{i_{eff}, V \}, x_u]
-\hbar \sum_s \hat \omega_{us} {\hat \delta {\hat {\bf V}} \over \hat \delta
x_s}
\rangle_{AV}~,~~~ \cr
}\eqno(G8)$$
giving the form of the second Ward identity analogous to Eq.~(G2).

We must now perform an important consistency check. If instead of taking
$V=H$ in the Ward identities of Eqs.~(G2) and (G8), we take instead
$V=(-1)^F H$, the effect is to simply interchange the roles of ${\bf H}$
and ${\bf \hat H}$ in the argument showing that the variation of the
equilibrium distribution does not contribute, and so this argument remains
valid.  Hence we get two new relations, which we must check are not in
contradiction with Eq.~(G3).  The two new relations are seen to be
identical when one uses the identity
$$\sum_s \hat \omega_{rs} {\delta {\bf H} \over \delta x_s}
=\sum_s \omega_{rs} {\hat \delta {\bf \hat H} \over \hat \delta x_s}
{}~~~,\eqno(G9a)$$
which like Eq.~(G6b) is a consequence of the fact that the same equations of
motion follow from the graded and ungraded trace variational principles.
When the index $u$ is bosonic, the new relation is easily seen to be
identical in form to Eq.~(G3), and so is automatically consistent.  When
the index $u$ is fermionic, we
introduce the definition $\tilde H_{eff}$ as in Eq.~(57a), and rewrite
Eq.~(G3) in the form
$$\langle [\tilde H_{eff}, x_u] \rangle_{AV}
=\hbar \langle \sum_s \omega_{us} {\delta {\bf H} \over \delta x_s}
\rangle_{AV} ~~~;
\eqno(G9b)$$
in this notation, the new relation takes the form
$$\langle \{\tilde H_{eff}, x_u\} \rangle_{AV}
=\hbar \langle \sum_s \hat \omega_{us} {\delta {\bf H} \over \delta x_s}
\rangle_{AV}~~~,
\eqno(G9c)$$
which involves a commutator rather than an anticommutator on the left hand
side. Treating separately the fermionic
cases in which $u=2r-1,~ x_u=q_r$ and $u=2r,~x_u=p_r $, and
taking linear combinations of Eqs.~(G9b) and (G9c) which eliminate
the right hand side, we find that the new relation of Eq.~(G9c) implies that
$$\eqalign{
\langle \tilde H_{eff} q_r \rangle_{AV} =&0~~~, \cr
\langle p_r \tilde H_{eff} \rangle_{AV} =&0~~~. \cr
}\eqno(G10)$$
These relations are compatible with what one expects for vacuum
expectations in quantum field theory, when one defines the fermionic vacuum
so that $\langle 0|p_{r~eff} = q_{r~eff} |0 \rangle =0$;
hence the new relation of Eq.~(G9c) is consistent
with the isomorphism conjectured in the text.  We also remark that Eq.~(G7c)
for the expectation of $\tilde F$, which can be rewritten as
$$\langle \sum_{r,F} q_r p_r \rangle_{AV} =0~~~, \eqno (G11a)$$
and Eq.~(G10) do not contradict the fermionic canonical algebra of Eq.~(60c),
because in general
$$q_r p_r \not = q_{r~eff} p_{r~eff}~~~;\eqno(G11b)$$
referring to the analysis of Appendix B we see that equality in Eq.~(G11b)
can hold only for the special case [see Eq.~(B5b)]
of complex Hilbert space with ${\cal E}=1$, which is thus ruled out.

Having established the equality of the ensemble averages of $\tilde C$ and
$\hat {\tilde C}$, and therefore a functional relationship between the
corresponding ensemble parameters $\tilde \lambda$ and
$\hat {\tilde \lambda}$, and $(-1)^F$, we can apply the argument of Eq.~(54c)
to conclude that the commutator terms involving $\tilde \lambda$ and
$\hat {\tilde \lambda}$ drop out of the Ward identities.  The Ward identities
of Eqs.~(G2) ad (G8) then simplify to
$$\eqalign{
0=&\langle
(\hat{\tau}(-1)^F\sum_{r} \alpha_{ur} \dot x_r
+\tau \dot x_u +\sum_s \omega_{us} \sigma_s \rho_s ) \hbar {\bf V} \cr
+& [{1 \over 2} \{i_{eff}, V \}, x_u]
- \hbar \sum_s \omega_{us} {\delta {\bf V} \over \delta x_s}
\rangle_{AV}~,~~~ \cr
}\eqno(G12a)$$
and
$$\eqalign{
0=&\langle
(\tau(-1)^F\sum_{r} \alpha_{ur} \dot x_r
+\hat{\tau} \dot x_u +(-1)^F\sum_s \hat \omega_{us} \sigma_s \rho_s )
\hbar {\bf \hat V} \cr
+& [{1 \over 2} \{i_{eff}, V \}, x_u]
-\hbar \sum_s \hat \omega_{us} {\hat \delta {\hat {\bf V}} \over \hat \delta
x_s}
\rangle_{AV}~,~~~ \cr
}\eqno(G12b)$$
with Eq.~(G12a) the basis of further applications as discussed in the text.
For any $V$ obeying the conditions
$$\eqalign{
\sum_s \omega_{us} {\delta {\bf V} \over \delta x_s}=&
\sum_s \hat \omega_{us} {\hat \delta {\bf {\hat V}} \over \hat \delta x_s}~~~,
\cr
\sum_s \hat \omega_{us} {\delta {\bf V} \over \delta x_s}=&
\sum_s \omega_{us} {\hat \delta {\bf {\hat V}} \over \hat \delta x_s}~~~, \cr
}\eqno(G13a)$$
an analysis paralleling the consistency check on the Hamiltonian given above
shows that the Ward identities obtained from Eqs.~(G12a, b) when the
$\hat \tau$ and $\tau$ terms are neglected, and those similarly obtained
when $V$ is replaced by $(-1)^F V$, are consistent as long as the conditions
$$\eqalign{
\langle \tilde V_{eff} q_r \rangle_{AV} =&0~~~, \cr
\langle p_r \tilde V_{eff} \rangle_{AV} =&0~~~, \cr
}\eqno(G13b)$$
with $\tilde V_{eff}={1 \over 2} \{ i_{eff},V \}$,
are obeyed for fermionic $r$.  These conditions are compatible with those
of Eq.~(G10), and have the same interpretation in terms of fermionic vacuum
structure.
\hfill \break
\parindent=25pt
\vfill\eject
\centerline{\bf References}
\bigskip
\noindent
[1] S. L. Adler, Nucl. Phys. B 145 (1994) 195 \hfill \break
\bigskip
\noindent
[2] S. L. Adler, Quaternionic quantum mechanics and quantum fields,
Secs.~13.5--13.7 and Appendix A (Oxford, New York, 1995) \hfill \break
\bigskip
\noindent
[3] S. L. Adler, G. V. Bhanot and J. D. Weckel, J. Math. Phys. 35 (1994)
531 \hfill \break
\bigskip
\noindent
[4] S. L. Adler and Y.--S. Wu, Phys. Rev. D 49 (1994) 6705 \hfill \break
\bigskip
\noindent
[5] H. Goldstein, Classical Mechanics, pp. 260--261
(Addison--Wesley, Reading MA, 1950)
\hfill \break
\bigskip
\noindent
[6]  D. ter Haar, Elements of Statistical Mechanics, Third ed., Sec. 5.13
(Butterworth Heinemann, Oxford and Boston, 1995)
\bigskip
\noindent
[7]  A. Sommerfeld, Thermodynamics and Statistical Mechanics, Secs.~28,
29, 36, and 40 (Academic Press, New York, 1956) \hfill\break
\bigskip
\noindent
[8] F. Mohling, Statistical mechanics: methods and applications, pp.
270--272  (Halsted Press/John Wiley, New York, 1982) \hfill \break
\bigskip
\noindent
[9] M. Kaku, Quantum Field Theory, pp. 407--410 (Oxford, New York, 1993)
\hfill\break
\bigskip
\noindent
[10] E. Witten, ``The 1/N Expansion in Atomic and Particle Physics'',
in Recent Developments in Gauge Theories, eds. G. 'tHooft et. al.
(Plenum Press, New York and London, 1980) \hfill\break
\bigskip
\noindent
[11] R. Gopakumar and D. J. Gross, ``Mastering the Master Field'', Princeton
preprint PUPT--1520, October, 1994 \hfill\break
\bigskip
\noindent
[12] I. Ya. Aref'eva and I. V. Volovich, ``The Master Field for QCD and
q--Deformed Quantum Field Theory'', Steklov preprint SMI--25--95,
November, 1995 \hfill\break
\vfill
\eject
\bye